\documentclass[]{aa}
\usepackage{natbib}
\usepackage[utf8]{inputenc}
\usepackage{graphicx}
\usepackage{subcaption}
\usepackage{txfonts}
\usepackage[dvipsnames]{xcolor}
\usepackage{amsmath}
\usepackage{comment}

\usepackage{hyperref}

\bibpunct{(}{)}{;}{a}{}{,} 

\newcommand{\unit}[1]{\,\mathrm{#1}}

\newcommand{\PaperII}{Paper~II}

\begin{document}
	\title{Modeling disks and magnetic outflows around a forming massive star: I. Investigating the two layer-structure of the accretion disk}
	\titlerunning{Magnetized accretion disks in massive star formation}
	\author{André~Oliva \inst{1}
	\and
	Rolf~Kuiper \inst{2,1}
	}

	\institute{Institute for Astronomy and Astrophysics, University of Tübingen, Auf der Morgenstelle 10, 72076, Tübingen, Germany\\
		\email{andree.oliva@uni-tuebingen.de}  \\
		\and
		Faculty of Physics, University of Duisburg--Essen, Lotharstraße 1, 47057, Duisburg, Germany\\
		\email{rolf.kuiper@uni-due.de}
	}


	\abstract{
	Like their lower mass siblings, massive protostars can be expected to: a) be surrounded by circumstellar disks and b) launch magnetically-driven jets and outflows. The disk formation and global evolution is thereby controlled by advection of angular momentum from large scales, the efficiency of magnetic braking and the resistivity of the medium, and the internal thermal and magnetic pressures of the disk.
	}{
	We determine the dominant physical mechanisms which shape the appearance of these circumstellar disks, their sizes and aspect ratios.
	}{
	We perform a series of 30 simulations of a massive star forming from the gravitational collapse of a molecular cloud threaded by an initially-uniform magnetic field, starting from different values for the mass of the cloud, its initial density and rotation profiles, its rotational energy content, the magnetic field strength, and the resistivity of the material. The gas and dust is modeled with the methods of resistive magnetohydrodynamics, also considering radiation transport of thermal emission and self-gravity. We check for the impact of spatial resolution in a dedicated convergence study.
	}{
	After the initial infall phase dominated by the gravitational collapse, an accretion disk is formed, shortly followed by the launching of magnetically-driven outflows. Two layers can be distinguished in the accretion disk: a thin layer, vertically supported by thermal pressure, and a thick layer, vertically supported by magnetic pressure. Both regions exhibit Keplerian-like rotation and grow outwards over time. We observe the effects of magnetic braking in the inner $\sim 50\,\mathrm{au}$ of the disk at late times in our fiducial case. The parameter study reveals that the size of the disk is mostly determined by the density and rotation profiles of the initial mass reservoir and not by the magnetic field strength. We find that the disk size and protostellar mass gain scale with the initial mass of the cloud. Magnetic pressure can slightly increase the size of the accretion disk, while magnetic braking is more relevant in the innermost parts of the disk as opposed to the outer disk. From the parameter study, we infer that multiple initial conditions for the onset of gravitational collapse are able to produce a given disk size and protostellar mass.
	}{}


	\maketitle

\section{Introduction}

Observations and numerical (radiation-)MHD models of the formation of massive protostars along with their accretion disks and collimated jets have made clear progress of the last decade. Even though there are several observational examples of massive protostellar jets \citep[see e.g.][]{Purser2016}, direct observations and estimations of the radii of their associated accretion disks are rarer. The source for the HH 80-81 radio jet, for example, has an accretion disk with an estimated radius of $950\text{--}1300\unit{au}$, and an inner radius of $\lesssim 170\unit{au}$, with the mass of the protostar in the range of $4\text{--}18\unit{M_\odot}$ \citep{Girart2017, Carrasco-Gonzalez2010}.

Recently, a new generation of multi-scale observations of the star-forming region IRAS 21078+5211 \citep{Moscadelli2021multi, Moscadelli2022} have revealed the existence of a disk-jet system around a protostar of $5.6\pm 2 \unit{M_\odot}$. Emission peaks from molecular rotational transitions of $\mathrm{CH_3CN}$ and $\mathrm{HC_3N}$ trace the kinematical footprint of a Keplerian-like accretion disk of around $400\unit{au}$ in diameter surrounding the protostar. For the first time, the direct study of the launching mechanism of the jet has been possible, thanks to water maser observations performed with Very Long Base Interferometry (VLBI) \citep{Moscadelli2022}, which traced individual streamlines in the jet and reached a resolution of $0.05\unit{au}$.

In \cite{Moscadelli2022}, we modeled the disk-jet system in IRAS 21078+5211 with a resistive-radiation-gravito-magnetohydrodynamical simulation of a forming massive star starting from the collapse of a cloud core. We followed the evolution of the protostar, the formation of the accretion disk, and the launching, acceleration, collimation, propagation and termination of the magnetically-driven outflows, which self-consistently form within the computational domain. Thanks to the use of an axisymmetrical grid, we were able to reach unprecedented resolutions in the regions surrounding the forming massive star (up to $0.03\unit{au}$), allowing for the comparison with the VLBI data (which has a resolution of $0.05\unit{au}$). We found good agreement between the simulated and observed disk size (given the protostellar mass) as well as the main features of the jet. The simulation was part of a larger study that examines the effects of the natal environment on the formation of a massive star. We present the full series of simulations in this article, with special focus on the dynamics of the accretion disk, while the dynamics of the magnetically-driven outflows are the focus of an upcoming article (hereafter Paper II).

Previously, the numerical simulations performed by \cite{Anders2018} were able to show the self-consistent magneto-centrifugal launching of the jet and a tower flow driven by magnetic pressure in the context of massive star formation. The present study builds on their work by improving the treatment of the thermodynamics of the gas and dust with the inclusion of radiation transport with the gray flux-limited diffusion approximation \citep{Kuiper2020}, and increasing the resolution of the grid. Those changes enable us to examine the vertical structure of the accretion disk. Previous numerical studies of the formation of a massive star in an environment with magnetic fields did not considered magnetic diffusion (e.g. \citealt{Myers2013} and \citealt{Seifried2011}) or approximated the thermodynamics of the system with the use of a barotropic \citep[e.g.][]{Machida2020} or isothermal \citep{Anders2018} equation of state instead of solving for radiation transport. Recent studies by \cite{Mignon-Risse2021} and \cite{Commercon2022}, which include ambipolar diffusion as the non-ideal MHD effect and do include radiation transport, find indications of a magneto-centrifugally launched jet (although their grid does not properly resolve the launching region), and smaller disks with stronger magnetic fields. The question of the size of the disk is explored in more detail in the present article, as we find an opposite trend in our results. For further literature review, we refer the reader to Sect. \ref{S: previous}.

Because of computational costs, previous efforts have focused on only a handful of natal environmental conditions for the massive (proto)star. Our setup, in turn, allows us to explore a wide range of natal environmental conditions and their impact on the outcome of the formed massive star and its disk-jet system.  In section \ref{S: model method} we describe the model, physical effects considered, and the parameter space of the simulations. We present an overview of the evolution of the system based on the fiducial case of the parameter space in \ref{S: evolution}. In section \ref{S: dynamics}, we focus on the dynamical processes observed in the accretion disk, i.e., the effect of the magnetic field, Ohmic dissipation and radiation transport. In section \ref{s:parameter-study} we explore the dependence of the disk size and evolution with the initial conditions for the gravitational collapse, and finally in section \ref{S: previous} we offer a comparison of our results with previous numerical studies.

\section{Model and method} \label{S: model method}
\subsection{Description of the model}
We model axisymmetrically the gravitational collapse of a cloud core of mass $M_C$ and a radius of $R_C=0.1\unit{pc}$ (cf. Fig. \ref{overview_evol}), which has an initial density profile of the form
\begin{equation}
\rho(r,t=0) = \rho_0 \left(\frac{r}{r_0}\right)^{\beta_\rho}.
\end{equation}
The constant $r_0$ is chosen to be $1\unit{au}$ and $\rho_0$ is determined by computing the mass of the cloud with the integral of the density profile over $0 < r < R_C$. The cloud is in slow initial rotation, given by an initial angular velocity profile of the form
\begin{equation}
\Omega(R,t=0) = \Omega_0 \left( \frac{R}{R_0} \right)^{\beta_\Omega},
\end{equation}
where $R$ is the cylindrical radius and $R_0$ is chosen to be $10\unit{au}$. The constant $\Omega_0$ is fixed by computing the ratio of rotational to gravitational energy $\zeta \equiv E_r/E_g$ from the initial density and rotation profiles; $\zeta$ is then a parameter of the simulation.

The cloud core is threaded by an initially uniform magnetic field directed parallel to the rotation axis. Its magnitude $B_0$ is determined by the normalized mass-to-flux ratio $\bar \mu$,
\begin{equation}
\bar \mu \equiv \frac{M_C/\Phi_B}{\left(M_C/\Phi_B\right)_\mathrm{crit}}.
\end{equation}
The denominator of the previous expression represents the critical value of the mass-to-flux ratio, i.e., the value for which the gravitational collapse is halted by the magnetic field under idealized conditions. We take the value from \cite{MouschoviasSpitzer1976}. The magnetic flux $\Phi_B$ is simply computed as the flux across the midplane, $B_0 \cdot \pi R_C^2$, which in magnitude is the same flux that enters or leaves the spherical cloud. 

\subsection{Physics}

The simulations solve for the equations of magnetohydrodynamics, with the addition of Ohmic resistivity as a non-ideal effect, self-gravity, the gravitational force from the forming massive star, and radiation transport for the thermal emission from the gas and dust, on an axisymmetric grid. Next, we describe the set of equations solved.

The dynamics of the gas and dust are followed by numerically solving the following system of conservation laws, using the Pluto code \citep{Mignone2007}:
\begin{equation}
\partial_t \rho + \vec \nabla \cdot (\rho \vec v) = 0,
\end{equation}
\begin{equation}
{\partial_t (\rho \vec v) } + \vec \nabla \cdot \left[  \rho \vec v \otimes \vec v -  \tfrac{1}{4\pi}{\vec B \otimes \vec B} + P_t  \mathsf{I} \right] = \rho \vec{a}_\text{ext},
\end{equation}
\begin{equation}
{\partial_t \vec B} + \vec \nabla \times (c \vec{\mathcal E}) = 0,
\end{equation}
\[ {\partial_t (E^K+E^\text{th}+E^B)} \]
\begin{equation} \label{e:eneq}
\quad\quad + \vec \nabla \cdot \left[ (E^K+E^\text{th} +  P)\vec v + c \vec{\mathcal E} \times \vec B \right] = \rho \vec v \cdot \vec a_\text{ext},
\end{equation}
which correspond to the continuity equation, momentum equation, induction equation and energy conservation equation, respectively. The gas is weakly ionized and has density $\rho$, velocity $\vec v$ and magnetic field $\vec B$. The magnetic field must also satisfy the solenoidality condition $\vec \nabla \cdot \vec B = 0$. The total pressure $P_t = P + \tfrac{1}{8\pi}{B^2}$ is composed of the thermal and magnetic pressures. The electric field $\vec{\mathcal E}$ can be directly substituted by the expression
\begin{equation}
c\vec{\mathcal E} = -\vec v \times \vec B + \eta \vec \nabla \times \vec B,
\end{equation}
where $\eta(\rho,T)$ is the Ohmic resistivity. The energy density of Eq. \ref{e:eneq} is separated into its components: kinetic ($E^K$), thermal or internal ($E^\mathrm{th}$) and magnetic ($E^B$). The acceleration source term $\vec a_\text{ext}$ is a sum of the gravity of the forming star $\vec a_{g\star} = -GM_\star/r^2\, \vec e_r$, the self-gravity of the gas $\vec a_\text{sg}$ and the viscosity source term $\vec a_\nu$.

The self-gravity acceleration source term is given by $\vec a_\text{sg} = -\nabla \Phi_\text{sg}$. The self-gravity potential $\Phi_\text{sg}$ is found by solving Poisson's equation $\nabla^2 \Phi_\text{sg} = 4\pi G\rho$ with a diffusion Ansatz, according to \cite{Kuiper2010circ}. Self-gravity is inherently a three-dimensional effect, and a cloud with the characteristics we consider here is expected to produce an accretion disk with spiral arms. Since we perform the simulations on a two-dimensional axisymmetric grid, we mimic the missing angular momentum transport by the gravitational torques with the addition of a shear viscosity source term $\vec a_\nu = \vec \nabla \Pi/\rho$, where the viscosity tensor $\Pi$ is fixed with the $\alpha$-parametrization of \cite{ShakuraSunyaev1973} and no bulk viscosity is considered. This approach was extensively studied in \cite{Kuiper2011}, and we refer the reader to that article for more details on the computation of the viscosity tensor, as well as comparisons of the efficacy of angular momentum transport using this approach in comparison to non-viscous 3D models.

We treat radiation transport with the gray flux-limited diffusion approximation, by using the module Makemake described in ample detail in \cite{Kuiper2020}. Specifically, we point  the reader to sections 2.3.2 and 2.3.4 in that reference. In a nutshell, two equations are solved simultaneously for treating radiation transport. The first one is the zeroth-moment of the radiation transport equation within a given volume, which essentially says that the net emitted energy density (emission minus absorption) that is not lost through the boundary as a radiative flux, has to be stored as a local radiative energy density. This equation is simplified by assuming  that the radiative flux can be written as a diffusion term in terms of the radiation energy density (hence flux-limited diffusion). The second equation describes the changes to $E^\text{th}$: the net absorbed energy density (absorption minus emission) has to be stored as internal energy in the gas. Both energy fields are solved without previous assumption of equilibrium between them; this is the so-called two-temperature approach \cite{Commercon2011rad}. In all but one of the simulations (see Sect. \ref{S: disk var eta} for details), a constant value of the opacity for the dust and gas of $1\unit{cm^2\, g^{-1}}$ was used. The initial dust-to-gas mass ratio is set to $1\%$. The irradiation from the forming massive star is not considered in this study.

We use the resistivity model by \cite{Machida2007} (which is based on a numerical study by \citealt{Nakano2002})
\begin{equation}
\eta = \frac{740}{X_e} \left({\frac{T}{10\unit{K}}}\right)^{1/2} \left[  1 - \tanh \left( \frac{n_H}{10^{15} \unit{cm^{-3}}} \right) \right] \unit{cm^2\, s^{-1}},
\end{equation}
where we use for simplicity $n_H=\rho/\mu_H$ as the number density of hydrogen nuclei ($\mu_H$ being the molecular weight of hydrogen) and the ionization fraction $X_e$ is given by
\begin{equation}
X_e = 5.7\cdot 10^{-4} \left(\frac{n_H}{\unit{cm^{-3}}}\right)^{-1}.
\end{equation}

At all moments in time, the properties of the formed massive star are computed using the evolutionary tracks for high-mass stars calculated by \cite{HosokawaOmukai2009evol} and the instantaneous values of the stellar mass and accretion rate.

\subsection{Boundary conditions}
The computational domain assumes symmetry with respect to the rotation axis and equatorial symmetry with respect to the midplane. This means that scalar quantities and the velocity are simply reflected across the $z$-axis and the midplane. In the case of the magnetic field, it is reflected across the $z$-axis, but across the midplane, the parallel component switches sign instead of the normal component. The same symmetries imply that the boundaries across the azimuthal direction are periodic. Both the inner and outer boundaries (i.e., the sink cell and the outermost part of the cloud) impose a zero gradient condition for the magnetic field and the azimuthal and polar components of the velocity; for the radial component of the velocity and the density, only outflow but no inflow is allowed. Therefore, all matter that goes inside of the sink cell is considered as accreted.

\subsection{Initial conditions: parameter space and numerical configuration}

\begin{table}
\caption{Grid resolutions}
\label{t:grid}
\centering
\begin{tabular}{c c c c c}
\hline\hline
Grid & $N_r$ & $N_\theta$ & $\Delta x_\text{min}\ [\unit{au}]$ & $\Delta x_{1000}\  [\unit{au}]$ \\
\hline
x1 & 56 & 10 & 0.51 & 160 \\
x2 & 112 & 20 & 0.25 & 80 \\
x4 & 224 & 40 & 0.12 & 40 \\
x8 & 448 & 80 & 0.06 & 20\\
x16 & 896 & 160 & 0.03 & 10\\
\hline 
\end{tabular}
\end{table}

We present an extensive parameter study of the system, with the express aim of aiding in the understanding of the physical processes that intervene in the formation and evolution of the disk and the magnetically-driven outflows. In total, we present results from 31 different runs. Instead of describing all the parameters in multiple tables, we show a summary of the parameter space as a diagram in the right panel of Fig. \ref{overview_evol}.

We used five different axisymmetrical grids with increasing resolution, that we name x1, x2, x4, x8 and x16. The grid for the radial coordinate $r$ increases logarithmically with distance, starting from the radius of the inner boundary ($3\unit{au}$ in the fiducial case), to the radius of the cloud core ($0.1\unit{pc}$). We often refer to the inner boundary as the sink cell. Using the assumed symmetries of the system, the colatitude $\theta$ extends from $0$ to $\pi/2$ radians, and it is divided linearly into cells such that they have approximately the same dimensions in both the radial and polar directions for each distance to the center of the cloud. Table \ref{t:grid} details the number of cells in each direction, the minimum cell size and a reference cell size at $1000\unit{au}$ for all the grids used in this study, and assuming the sink cell size of the fiducial case.

The parameters with direct relation to physical quantities we considered are: the mass of the cloud core $M_C$, the resistivity model used, the normalized mass-to-flux ratio $\bar \mu$, the initial density profile exponent $\beta_\rho$, the initial rotation profile exponent $\beta_\Omega$, and the ratio of the rotational-to-gravitational energy $\zeta$. The values for each parameter in the fiducial case are highlighted in blue in Fig. \ref{overview_evol}. The fiducial case was run on all the grids. Modifications of the fiducial case are then investigated while keeping the rest of the parameters constant. Because of this, we refer in the paper to each simulation by the modified parameter followed by the grid it was run on; for example, $M_C = 150\unit{M_\odot}\text{ [x8]}$ refers to a cloud of $150\unit{M_\odot}$, with the rest of the parameters as in the fiducial case for grid x8. The full set of values with direct relation to physical quantities was run on the base grid x4, and some selected values, which are underlined in Fig. \ref{overview_evol}, were run on grid x8 as well. During the preparation of this study, we also ran a part of this physical parameter space on the lower resolution grids, however, we do not include the results here because they are redundant. Details of the choice of the parameters, as well as the comparison of results obtained, are presented in Sect. \ref{s:parameter-study}.

The numerical convergence of the results was studied by changing the values of the sink cell radius, the strength of the $\alpha$ shear viscosity and the Alfvén limiter using the grid x2 as a base. A more detailed explanation of these latter  parameters, as well as the results of the convergence study can be found in Appendix \ref{s:convergence}.


\section{Overview of the evolution of the system}\label{S: evolution}

\begin{figure*}
	\includegraphics[width=\textwidth]{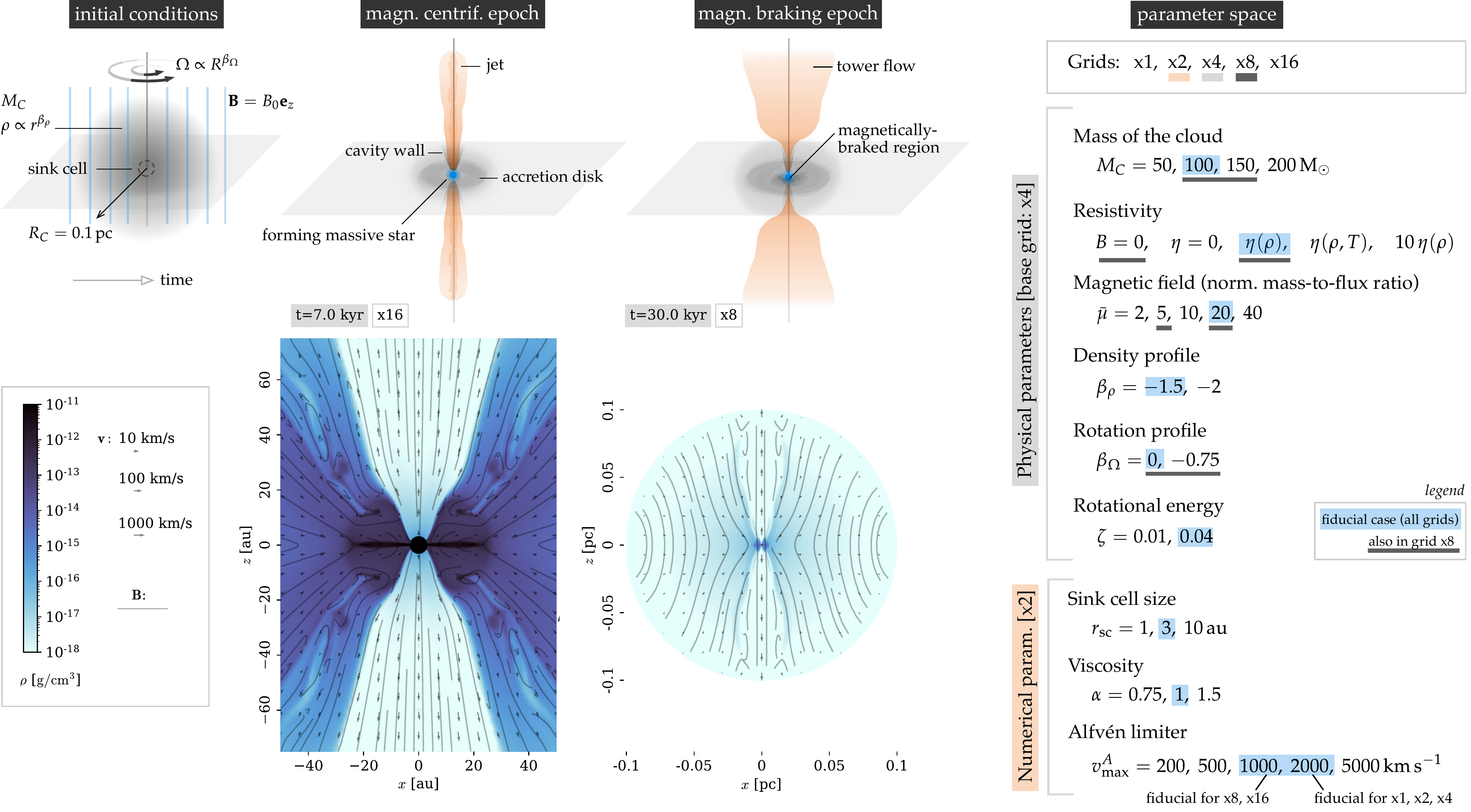}
	\caption{Initial conditions, overview of the evolution of the system, and summary of the parameter space presented in this study.}
	\label{overview_evol}
\end{figure*}

Figure \ref{overview_evol} shows a schematic overview of the features observed during the evolution of the system, based on the results of the fiducial case but generally present in most of the simulations of the study. We describe these characteristic features and processes in the following subsections case-by-case.

\subsection{Gravitational infall epoch}
As soon as the simulation begins, the cloud starts collapsing under its own gravity. The flow quickly becomes super-Alfvénic (the kinetic flow velocity is larger than the Alfvén speed), and the magnetic field lines are dragged by the infall accordingly. The result of this collapse is that the configuration of magnetic field lines adopts an ``hourglass'' shape, as described in \cite{Galli2006}, and that has been observed in \cite{Beltran2019}.

\subsection{Protostar}
A massive (proto)star is formed at the center of the cloud, which reaches a mass of $\sim 20\unit{M_\odot}$ after $30\unit{kyr}$. The luminosity of the (proto)star is dominated by the conversion of gravitational energy into radiation during the first $20\unit{kyr}$, and only after that time the stellar luminosity becomes comparable to the accretion luminosity. We point out that both the stellar and accretion luminosities are not taken into account in this study, but we will report on these radiative feedback effects in future work.

\subsection{Magneto-centrifugal epoch}

Due to the conservation of angular momentum, gas coming from large scales rotates faster as it reaches the center of the cloud. At $t\sim 5\unit{kyr}$, enough angular momentum is transported onto the center of the cloud to form a Keplerian-like accretion disk as a result. The disk grows in time due to the continued infall of material from large scales.

The accretion disk can be morphologically divided into a thin and a thick layer. The thin layer of the disk  is a region of the midplane consisting of material in rotation with speeds close to the Keplerian speed. The density of the thin layer increases towards the center of the cloud $10^{-16}$ to $10^{-11}\unit{g\, cm^{-3}}$. Enclosing the thin layer, there is a less dense thick layer (typical densities: $10^{-14}\unit{g\,cm^{-3}}$) which is also rotating.

Roughly at the same time as the disk forms, magnetically-driven outflows are launched, which we classify according to their speeds and driving mechanisms. The jet is launched by the magneto-centrifugal mechanism (in a way similar to the model of \citealt{BlandfordPayne1982}) and it reaches typical speeds higher than $100\unit{km\,s^{-1}}$. Rotation drags the magnetic field lines, as it has been observed in \cite{Beuther2020} and \cite{Girart2013}. This creates a magnetic pressure gradient that drives a slower and broader tower flow (cf. the model of \citealt{Lynden-Bell2003}). Its typical speeds are of the order of $10\unit{km\,s^{-1}}$, and it broadens with time. A discussion of the dynamical processes present in the launching of outflows is offered in \PaperII.

On the interface between inflow along the thick layer of the disk and outflow along the cavity, an additional structure is identified, which we refer as the \emph{cavity wall}. Densities at the cavity wall are higher than those in the cavity, but also higher than the surrounding infalling envelope and the thick layer of the disk. In terms of its dynamics, material from the cavity wall contributes episodically to both the inflow and the outflow.

\subsection{Magnetic braking epoch}

As magnetic field lines are dragged by rotation, the resulting magnetic tension exerts a torque on the gas that brakes it, infalling as a consequence of the angular momentum loss. The inclusion of Ohmic dissipation makes magnetic braking negligible at early times in the simulation, but at around $t\sim 15 \unit{kyr}$, magnetic braking starts to affect the innermost region of the disk and the cavity wall ($r\lesssim 30\unit{au}$). As a result, a magnetically-braked region is formed, where the material is mostly infalling but where the disk densities are kept because of constant replenishment through material from the cavity wall.

The extraction of angular momentum from the inner region of the cloud by magnetic braking modifies or even interrupts the magneto-centrifugal mechanism that drives the jet, but not the tower flow, which is mostly present at large scales. The tower flow broadens over time because the continuous dragging of magnetic field lines by rotation increases the magnetic pressure gradient on top of the disk, which is itself growing in time as well. This outflow broadening mechanism offers an explanation for the earliest outflow broadening observed and discussed in \cite{Beuther2005}. An analysis and discussion of the effects of magnetic braking in outflows, as well as their propagation over time into the large scales of the cloud is offered in \PaperII, which focuses on the outflow physics of the same simulation data analyzed here.

\section{Dynamics of the disk} \label{S: dynamics}

\begin{figure}
	\centering
	\includegraphics[width=\columnwidth]{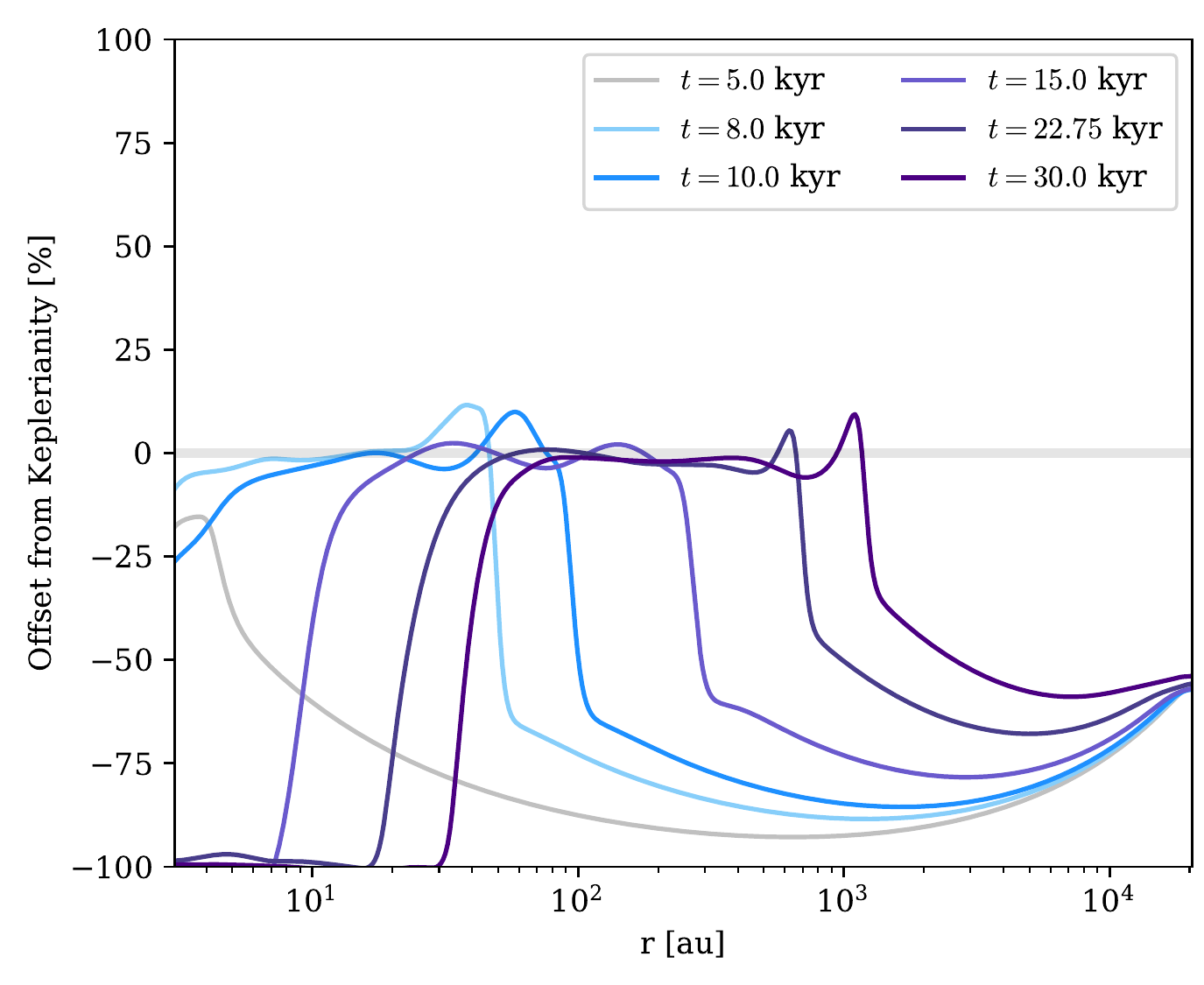}
	\caption{Keplerianity on the midplane, computed with the fiducial simulation on grid x8.}
	\label{thindisk_keplerianity}
\end{figure}

\begin{figure*}
	\centering
	\includegraphics[width=\textwidth]{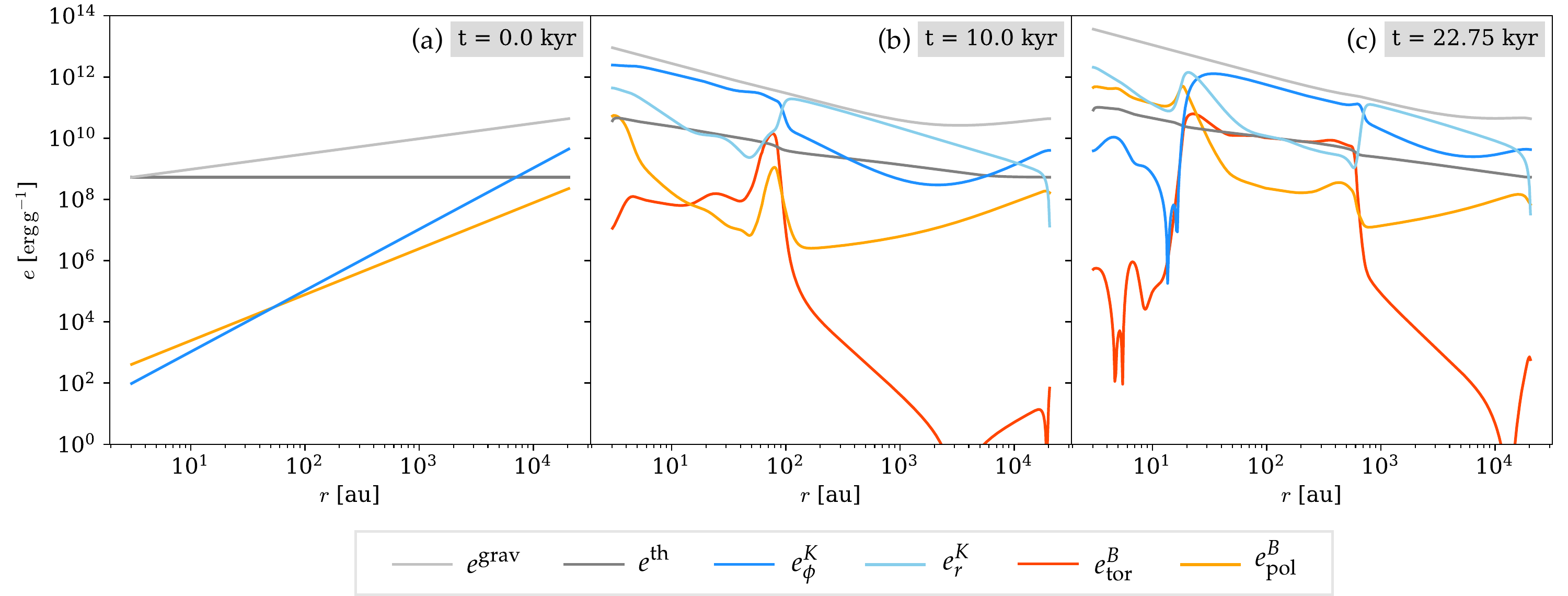}
	\caption{Contributions to the specific energy in the thin layer of the disk, calculated with the fiducial simulation on grid x8.}
	\label{thindisk_energies}
\end{figure*}

\begin{figure*}
	\centering
	\includegraphics[width=\textwidth]{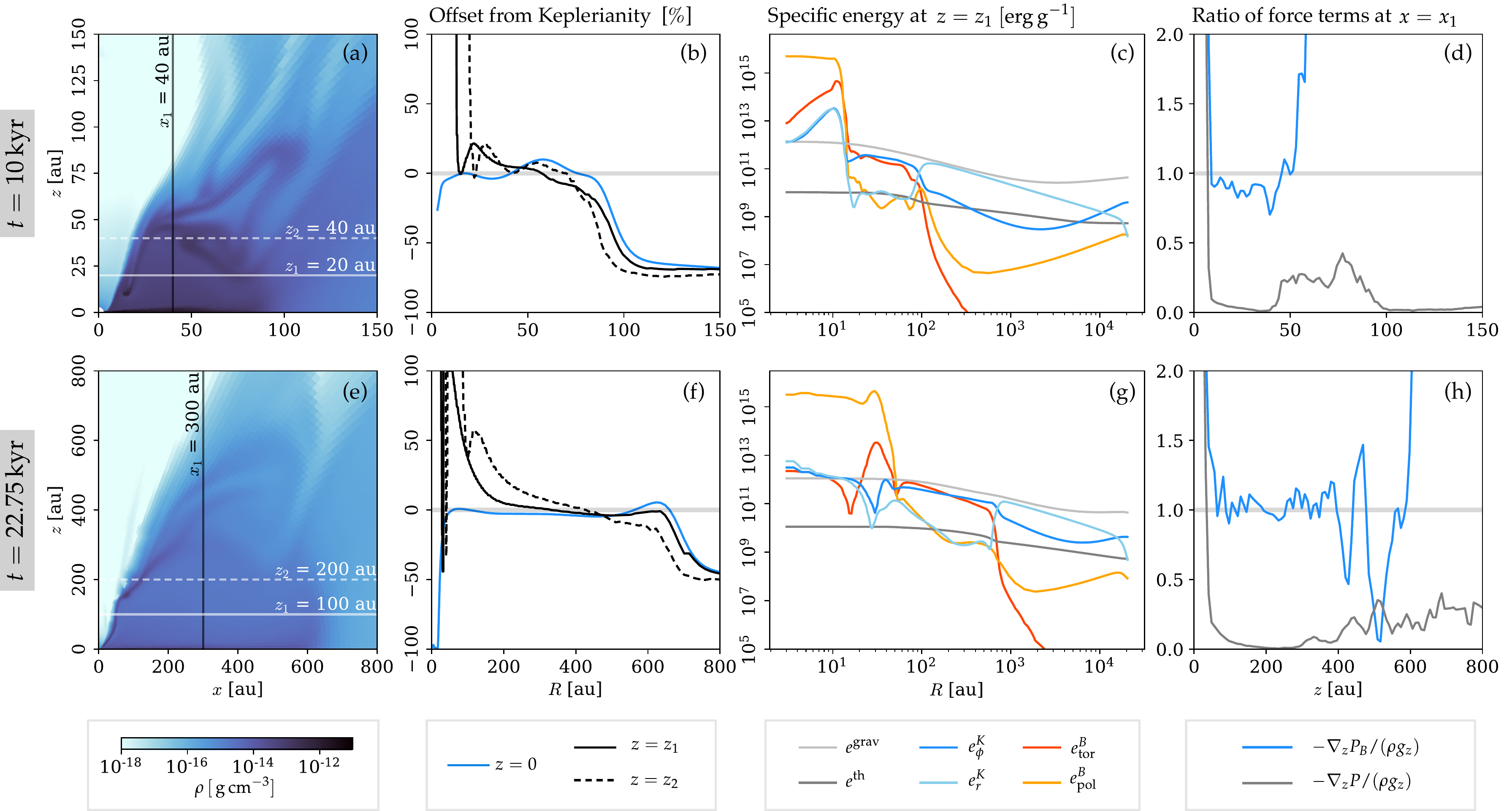}
	\caption{Dynamics of the thick layer of the disk. The panels of the figure are grouped in columns such that they show: the density structure of the thick layer (\emph{a} and \emph{e}), the offset from Keplerianity at three different heights in the thick layer (\emph{b} and \emph{f}), the contributions to the specific energy (\emph{c} and \emph{g}), and the vertical balance of forces (\emph{d} and \emph{h}). Data from the fiducial simulation on grid x8 was used for this figure.}
	\label{thickdisk_dynamics}
\end{figure*}

After giving an overview on the general evolution of the system, we aim at understanding the dynamical processes present in the disk in detail by taking the fiducial case and examining selected terms of the equations of motion. With the identification of these processes, we are able to perform a more effective and comprehensive comparison of results for the full parameter space, which we present in Sect. \ref{s:parameter-study}.

\subsection{Radial dynamics in the disk}

The forming massive star accretes mass through the accretion disk, which implies that the velocity in the disk must have a component in the radially-inwards direction. The torque necessary for the angular momentum transport in the disk is provided by gravitational torques of forming spiral arms, which is modeled here through the use of the $\alpha$ shear viscosity model. Nevertheless, $v_\phi \gg v_R$ in the disk, and so, it is useful to examine the effects of the different forces that act in the disk under the assumption of radial equilibrium.

In cylindrical coordinates $(R,\phi,z)$, the radial component of the MHD momentum equation, setting $\theta=\pi/2$ (midplane), assuming axisymmetry, is given by
\begin{equation}\label{e:mhd-R}
\frac{\partial (\rho v_R)}{\partial t} + \nabla \cdot \left( \rho v_R \vec v - \frac{B_R \vec B}{4\pi}  \right) + \frac{\partial P_t}{\partial R} = \rho g_R + \frac{\rho v_\phi^2 - B_\phi^2/(4\pi)}{R},
\end{equation}
where the divergence operator $\nabla \cdot$ acts according to
\begin{equation}\label{e:nablaop}
\nabla \cdot \vec f(R,z) = \frac{1}{R}\frac{\partial (R f_R)}{\partial R} + \frac{\partial f_z}{\partial z}
\end{equation}
onto the vector fields $\vec f(R,z) \in \{ \rho v_R \vec v, B_R\vec B\}$; $\vec g$ is the total gravitational force from the star and self-gravity, and $P_t$ is the sum of the thermal and magnetic pressures.

In order to examine the forces that define radial equilibrium in the disk, we consider a parcel of gas moving in a circular orbit, for which we set $v_R = 0$ in eq. \ref{e:mhd-R} and obtain

\begin{equation}\label{e:mhd-R2}
 \frac{\rho v_\phi^2 - B_\phi^2/(4\pi)}{R} = -\rho g_R + \frac{\partial P}{\partial R} +  \frac{\partial}{\partial R}\left(\frac{B^2}{8\pi}\right) -\nabla \cdot  (B_R \vec B).
\end{equation}

For $\vec B = \vec 0$, $P = 0$, and considering the midplane, one recovers the well-known condition for gravito-centrifugal equilibrium (on the co-rotating frame), which defines the Keplerian velocity, $v_K = \sqrt{GM(r)/R}$; $M(r)$ being the enclosed mass (dominated by the mass of the protostar for small $r$). The use of the enclosed mass is an approximation to the use of the full potential given by the Poisson equation.

\subsubsection{Keplerianity}
The degree of gravito-centrifugal equilibrium in the whole disk can be measured by computing the ratio of the centrifugal force density to the gravitational force density:
\begin{equation}
\frac{a^\mathrm{c}_R}{\rho g_R} = \frac{\rho v_\phi^2}{r\sin\theta} \cdot \frac{r^2}{\rho G M(r) \sin\theta} = \frac{v_\phi^2}{v_K^2\sin^3\theta} .
\end{equation}
This motivates us to define the offset from Keplerianity as
\begin{equation}
\text{offset from Keplerianity} = \left|\frac{v_\phi}{v_K\sin^{3/2}\theta}\right|-1,
\end{equation}
which in the midplane corresponds to the fractional difference of the azimuthal velocity with respect to the Keplerian value.

Figs. \ref{thindisk_keplerianity} and \ref{thickdisk_dynamics} show the offset from Keplerianity presented as a percentage, in such a way that the negative values indicate sub-Keplerianity, and the positive values, super-Keplerianity. In the midplane, there is a Keplerian region within $\pm 25 \%$ for all times, which corresponds to the thin accretion disk. The Keplerianity drops outside of the disk, which corresponds to the infalling envelope. At the exterior boundary of the disk, the mostly-rotating material encounters the mostly-infalling material from the region of the envelope, creating a centrifugal barrier that is seen as a slight super-Keplerianity in Fig. \ref{thindisk_keplerianity}. Even though we ignore the effect of viscosity in the analysis of this section, we note that, as it transports angular momentum outwards, its effect is to increase positively the deviation from Keplerianity. The rest of the deviations from Keplerianity in the disk are discussed in the following sections.

Panels \emph{b} and \emph{f} of Fig. \ref{thickdisk_dynamics} contrast the values of the offset from Keplerianity in the midplane with values in planes parallel to it, which slice through the thick layer of the disk and reveal its dynamical structure. In general, the thick layer also shows Keplerianity within $\pm 25\%$, but it decreases with altitude because the cylindrical-radial component of gravity also decreases with altitude. The regions close to the cavity become increasingly super-Keplerian, while the regions close to the infalling envelope become increasingly sub-Keplerian.

\subsubsection{Specific energies}
We study the dynamics of the thin and thick layers of the disk by computing the terms of the Bernoulli equation, that is, the contributions to the specific energy of the material. This allows us to isolate the role of each term in the equations of magnetohydrodynamics and compare its relative importance. The results of the energy calculations are available in Fig. \ref{thindisk_energies} for the midplane, and Fig. \ref{thickdisk_dynamics} for the plane $z=z_1$. The contributions to the specific energy we considered are:
\begin{itemize}
\item the gravitational specific energy \begin{equation}e^\text{grav} = \frac{GM(r)}{r},\end{equation}
\item the thermal specific energy \begin{equation} \label{e:energies-thermal}
e^\text{th} = \frac{P}{\rho (\Gamma - 1)},
\end{equation}
\item the contribution to the specific kinetic energy by the azimuthal component of velocity \begin{equation}
e^{K}_\phi = \frac{v_\phi^2}{2},
\end{equation}
\item the contribution to the specific kinetic energy by the spherical-radial component of velocity
\begin{equation}
e^K_r = \frac{v_r^2}{2},
\end{equation}
\item the contribution to the specific magnetic energy by the toroidal component of the magnetic field
\begin{equation}
e^B_\text{tor} = \frac{B_\phi^2}{8\pi \rho},
\end{equation}
\item the contribution to the specific magnetic energy by the poloidal component of the magnetic field
\begin{equation}
e^B_\text{pol} = \frac{B_r^2 + B_\theta^2}{8\pi \rho}.
\end{equation}
\end{itemize}

Initially, the specific gravitational energy increases as a function of $r$ because it is given by the enclosed mass (see Fig. \ref{thindisk_energies}a). As the collapse progresses (Fig. \ref{thindisk_energies}b), the gravity of the forming massive star becomes more important, which means that the curve for $e^\text{grav}$ has a point-source-like behavior close to the center of the cloud. At large scales, the enclosed mass of the envelope becomes important. As expected, the infalling envelope has $e^K_r > e^K_\phi$, but the advection of angular momentum from large to small scales causes $e^K_\phi$ to increase until $e^K_\phi > e^K_r$, forming the disk. When the magnetic field lines are dragged by rotation in the disk, the toroidal contribution to the magnetic energy increases.

\subsubsection{Thermal and magnetic pressure} \label{S: dynamics - PB}

The following two sections examine the contributions of the remaining terms of eq. \ref{e:mhd-R2} to the equilibrium of cylindrical radial forces in the disk and their impact to the values of Keplerianity shown in Fig. \ref{thindisk_keplerianity}.

First, we only consider the additional radial support that the thermal pressure gradient $\partial P/\partial R$ gives to the material in the disk, which be inferred by using eq. \ref{e:energies-thermal} and Figs. \ref{thindisk_energies} and \ref{thindisk_denstemp}. In agreement with the theory, we find that the thermal pressure gradient provides an additional small support against gravity and therefore makes material in radial equilibrium to appear to be slightly sub-Keplerian.

In Gaussian units, the magnetic pressure and magnetic energy density correspond to the same quantity, which means that we can use the specific energy density to study the magnetic pressure support in the disk. In the infalling envelope, the poloidal component of the magnetic field contributes the most to magnetic pressure (see, e.g., Fig. \ref{thindisk_energies}b), given that the initial magnetic field distribution is parallel to the rotation axis. When the disk is formed, the toroidal component of the magnetic field increases in general because the magnetic field lines are dragged by rotation. The toroidal magnetic field becomes dominant over the poloidal component, which additionally allows us to neglect the magnetic tension term $\nabla \cdot (B_R \vec B)$ from eq. \ref{e:mhd-R2} in most parts of the disk.

The magnetic pressure $P_B = \rho (e^B_\text{tor} + e^B_\text{pol})$ decreases with distance in the infalling envelope (compare Figs. \ref{thindisk_energies}b and  \ref{thindisk_denstemp}a), and overall in the disk as well, although the more complex structure of the magnetic field in the thin and thick layers of the disk means that there are regions of local increase with distance. Similarly to our argument with the thermal pressure gradient, $v_\phi < v_K$ for a negative magnetic pressure gradient. Therefore, the radius of the disk should be slightly larger compared to the non-magnetic case, given that $v_\phi$ decreases with distance and the angular momentum contained at larger $R$ becomes sufficient to support a circular orbit. This argument is revisited in Sect. \ref{S: disk var B} after considering the effects of magnetic braking (Sect. \ref{S: disk MB}) and magnetic diffusion (Sect. \ref{S: disk m diff}).

Additionally to magnetic pressure, the second term of the left hand side of eq. \ref{e:mhd-R2} reveals that the azimuthal component of the magnetic field reduces the effectiveness of the centripetal force required to keep a circular orbit. This can be measured by comparing the azimuthal component of the kinetic and magnetic energies:
\begin{equation}
\frac{B_\phi^2}{4\pi R} \cdot \frac{R}{\rho v_\phi^2}  = \frac{e^B_\text{tor}}{e^K_\phi}.
\end{equation}
In the thin layer of the disk, the contribution of this term is negligible, but in the thick layer, they are comparable. In terms of Keplerianity, this term can cause the material to be slightly super-Keplerian; this can be observed when comparing panels \emph{f} and \emph{g} of Fig.  \ref{thickdisk_dynamics}.

\subsection{Ohmic dissipation} \label{S: disk m diff}
\begin{figure}
\includegraphics[width=\columnwidth]{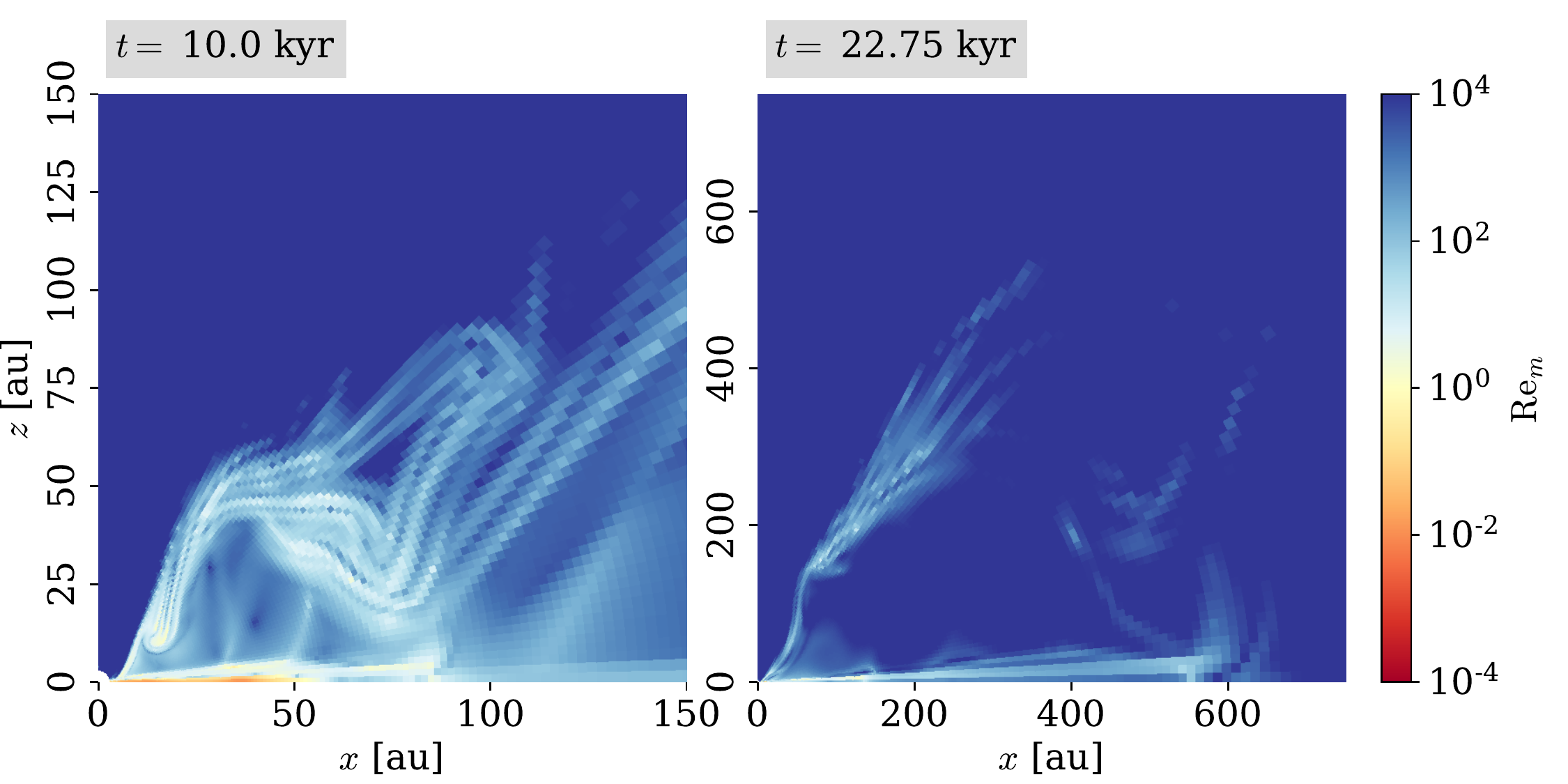}
\caption{Magnetic Reynolds number, computed with the fiducial simulation on grid x8. For $t=10\unit{kyr}$, the radius of the disk is around $80\unit{au}$, while for $t=22.75\unit{kyr}$, the exterior radius is $\approx 700\unit{au}$ and the interior radius is $\approx 20\unit{au}$.}
\label{Rem}
\end{figure}

According to the model of Ohmic resistivity by \cite{Machida2007}, which we use, the resistivity increases progressively with density for the number densities we obtain here ($n_H \lesssim 10^{13}\unit{cm^{-3}}$). We computed the ratio of the magnitude of the induction and diffusion terms in the induction equation, which is analogous to the magnetic Reynolds number, as
\begin{equation}\label{e:Rem}
\mathrm{Re}_m = \frac{|\vec \nabla \times (\vec v \times \vec B)|}{|\eta \vec \nabla \times  (\vec \nabla \times \vec B)|},
\end{equation}
and show the results in Fig. \ref{Rem}, which are computed for two instants of time from the fiducial case. At $t=10\unit{kyr}$, the plasma is fully diffusive in most regions of the thin layer of the disk; the thick layer is dominated by advection but it experiences magnetic diffusion in around one part per hundred; and the cavity wall is also partially diffusive. Later in time, at $t=22.75\unit{kyr}$, only the highest density regions close to the massive protostar are dominated by diffusion, the thin layer experiences magnetic diffusion of around one part per hundred, and the thick layer is almost completely dominated by advection.

Ohmic resistivity acts mostly on the toroidal component of the magnetic field, as it can be seen by comparing the energy curves in Figs. \ref{thindisk_energies} and \ref{thickdisk_dynamics}. In the thick layer of the disk at late times (Fig. \ref{thickdisk_dynamics}g) $e_\text{tor}^B$ roughly follows $e_\phi^K$. This sets the picture for the case where magnetic diffusion is low. On the other hand, in the thin layer of the disk, at early times (Fig. \ref{thindisk_energies}b), $e_\text{tor}^B$ and $e_\phi^K$ are clearly decoupled: the toroidal magnetic field component becomes low inside of the disk, corresponding to the region where diffusion dominates over advection. In the outer disk, where the density is lower, the resistivity is lower as well, and the toroidal component of the magnetic field is allowed to increase.

\subsection{Magnetic braking} \label{S: disk MB}
The momentum equation in the azimuthal direction, considering axisymmetry and circular motion, reduces to
\begin{equation}
\frac{\partial (\rho v_\phi)}{\partial t}  =  \frac{1}{R^2}\frac{\partial}{\partial R}(R^2 B_\phi B_R) + \frac{\partial}{\partial z}(B_\phi B_z).
\end{equation}
The right hand side of this equation is equal to zero if there are no magnetic fields, and it leads to the conservation of angular momentum. The terms on the right hand side constitute the azimuthal component of the magnetic tension force, which exerts a torque that reduces angular momentum locally. This process is known as magnetic braking, and it can be understood as a consequence of the dragging of magnetic field lines by rotation \citep{Galli2006}.

Even though Ohmic resistivity reduces the toroidal component of the magnetic field (cf. $e^B_\text{tor}$ in Fig. \ref{thindisk_energies}b), it does not completely suppress it. Over time, the magnetic field lines are wound enough so that magnetic braking happens in the innermost parts of the disk. In Fig. \ref{thindisk_energies}c, which corresponds to $t=22.75\unit{kyr}$,  $e^B_\text{tor}$ is higher in the disk than when it is measured at $t=10\unit{kyr}$, while $e^K_\phi$ decreases over the same interval of time in the inner $20\unit{au}$; those quantities show the reduction of angular momentum due to magnetic tension. In the face of the reduction of angular momentum, the material falls toward the protostar, as evidenced by the increase in $e^K_r$; this creates the magnetically-braked region. The cavity wall and the thick layer of the disk are also affected by magnetic braking; material from several directions is then delivered to the magnetically-braked region.

\subsection{Vertical support}

In the vertical direction (i.e., parallel to the rotation axis), the momentum equation reads
\begin{equation}\label{e:mhd-z}
\frac{\partial (\rho v_z)}{\partial t} + \nabla \cdot \left(\rho v_z \vec v  - \frac{B_z \vec B}{4\pi}\right) + \frac{\partial P_t}{\partial z} = \rho g_z
\end{equation}
with the same definition of the divergence as given in eq. \ref{e:nablaop}. Considering a parcel of gas moving in a circular orbit ($v_z=0$), and given that in the disk region $B_z \ll B_\phi$, eq. \ref{e:mhd-z} becomes
\begin{equation}
\frac{\partial P}{\partial z} + \frac{\partial}{\partial z} \left( \frac{B^2}{8\pi} \right) = \rho g_z.
\end{equation}

The ratio of the  thermal to magnetic pressure is equivalent to the ratio of the specific thermal energy to the total magnetic energy. In the thin layer of the disk (Fig. \ref{thindisk_energies}), thermal pressure dominates over magnetic pressure, from which we conclude that the thin layer is supported mainly by thermal pressure (at late times, magnetic pressure becomes comparable but not much higher than thermal pressure). On the contrary, the thick layer of the disk is supported by magnetic pressure, as evidenced by Figs. \ref{thickdisk_dynamics}d and \ref{thickdisk_dynamics}h, where we present a direct comparison of both pressure gradients to gravity in a vertical slice that crosses the disk. As the vertical component of gravity vanishes at the midplane, both curves increase towards $z=0$. In the thick layer of the disk, thermal pressure decreases, while the magnetic pressure gradient compensates gravity. At the cavity wall, thermal pressure increases because it has a higher density than the thick layer of the disk. Finally, in the cavity, magnetic pressure dominates over gravity in the vertical direction, which is expected in the case of a magnetically-driven outflow.

\subsection{Effects of other forms of magnetic diffusion} \label{S: other magn diff}
\begin{figure}
\centering
\includegraphics[width=0.6\columnwidth]{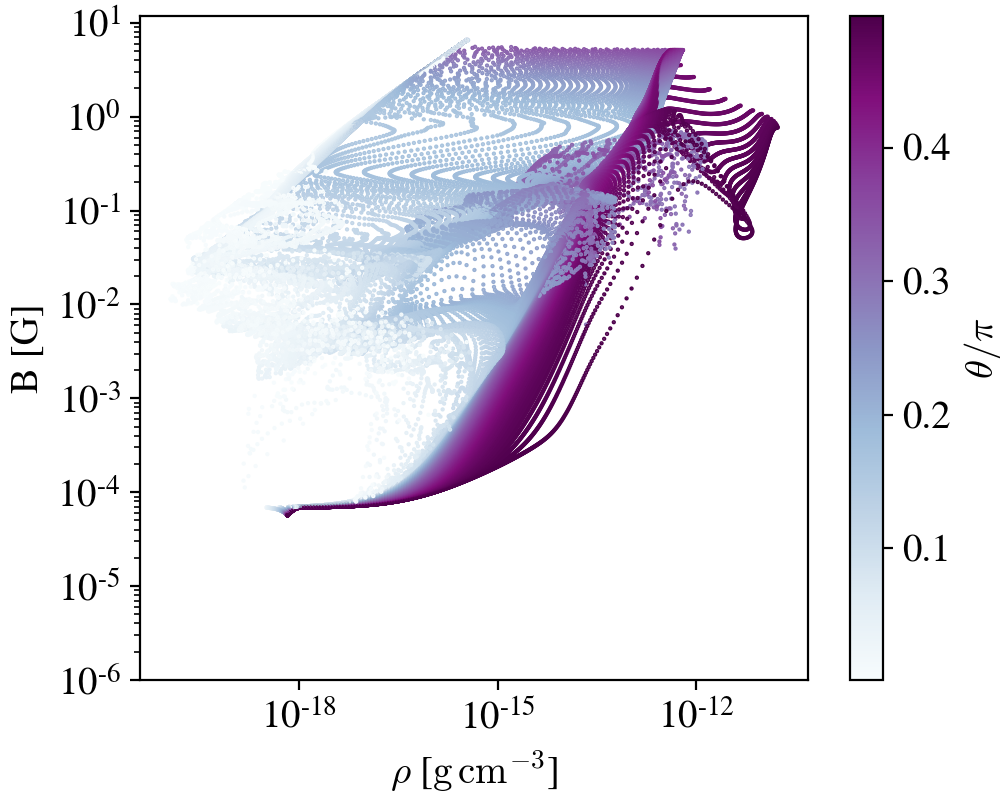}
\caption{Distribution of the magnetic field strength as a function of density. This histogram was computed with the fiducial simulation run on grid x16, for time $t=7\unit{kyr}$. The color scale indicates the polar angle: the dark purple colors correspond to the midplane (with the disk corresponding to $\rho \gtrsim 10^{-15} \unit{g\,cm^{-3}}$) and the light blue colors correspond to the rotation axis (where the jet is located).}
\label{brho}
\end{figure}

In this study, the effects of ambipolar diffusion and the Hall effect are not considered, and this formally constitutes a caveat of our approach which we plan to address in a future study. In this section, we discuss the expected effects that other forms of magnetic diffusivity might have on the results reported in the previous sections, with a focus on the fiducial case. We examine every region of interest for the discussion: the accretion disk, the infalling envelope and the outflow cavity. The simulation catalog also includes an investigation of the effects of the resistivity model; the results can be used to further assess the missing magnetic diffusivity and are presented in Sect. \ref{S: disk var eta}.

First, we consider only the accretion disk. The curves for ambipolar diffusivity and Ohmic resistivity given in e.g. \cite{Marchand2022,Marchand2016} and \cite{Tsukamoto2020} reveal that, at least for the range of number densities $10^8 \lesssim n_H \lesssim 10^{13}\unit{cm^{-3}}$, which are found in the thin layer of the accretion disk, the values of the diffusivities are similar within the uncertainties of the dust models considered for their derivation, differing by about a factor of ten. The distribution of the magnetic field strength with density is presented in Fig. \ref{brho}, where we use a color scale that highlights with darker colors the midplane. For the densities of the accretion disk ($\rho > 10^{-14}\unit{g\,cm^{-3}}$), the magnetic field strength increases with density, until it reaches between $0.1$ and $1\unit{G}$. When comparing the magnetic field distribution to the results of a recent study by \cite{Commercon2022}, who considered the effects of ambipolar diffusion but not Ohmic dissipation, we find consistent results for the thin layer of the disk. In the thick layer, however, the addition of higher, more realistic magnetic diffusivities might reduce the vertical magnetic pressure support. This means that thinner accretion disks than those reported here would be expected in a more realistic setting.

As for the effects of magnetic braking, we expect that ambipolar diffusion and the Hall effect reduce the magnetic Reynolds number, that is, the material would become more diffusive. Contrary to Ohmic dissipation, ambipolar and Hall diffusivities depend on magnetic field, and so the reduction could be of particular importance at later times, when considerable magnetic flux has been transferred to the central regions of the cloud core. This would cause a delay in the formation of the magnetically-braked region in the inner disk (see also the discussion in Sect. \ref{S: disk var eta}). On the other hand, magnetic braking could play an important role in determining the total angular momentum transferred to the forming massive star, as well as the growth of HII regions by terminating the magnetically-driven outflows (Martini et al., in prep.). An investigation of the role of all forms of magnetic diffusivity in the long-term evolution of the accretion disk and the jet is necessary. However, at least for early times ($t\lesssim 20\unit{kyr}$) in the formation of the massive star, our results fit observational constraints (see Sect. \ref{S: obs}).

As a second region of interest, we consider the infalling envelope. According to the curves by \cite{Marchand2022}, ambipolar diffusivities tend to be higher at lower densities ($n_H < 10^{8} \unit{cm^{-3}}$). This suggests that we are missing key magnetic diffusion during the early phases of gravitational collapse, where densities are low. However, there are theoretical and observational results that suggest that the collapse phase is still reasonably modeled in our simulations. In this simulation series, we consider the case of weak magnetic fields (see the discussion on this assumption in Sect. \ref{S: disk var B}). Because ambipolar diffusivities increase with magnetic field strength, we expect magnetic diffusion to play a stronger role in cases when the magnetic fields are initially strong. In the studies by \cite{Commercon2022} and \cite{Mignon-Risse2021}, who considered $\bar \mu = 5$ and ambipolar diffusion but no Ohmic dissipation, the magnetic field distribution for low densities coincides well with what is depicted in Fig. \ref{brho}. Based on this, we would not expect a strong qualitative divergence of our results for the gravitational collapse epoch ($t \lesssim 5 \unit{kyr}$) upon the addition of ambipolar diffusion to our models. From the observational point of view, the role of ambipolar diffusion as a necessary enabler for gravitational collapse in a magnetized cloud core was recently questioned in \cite{2022Natur.601...49C}. These latter authors found a supercritical low-mass prestellar core, which is not yet self-gravitating. This means that the magnetic field was diffused before the gravitational collapse takes place, that is, earlier than predicted by the classical picture of ambipolar diffusion turning a subcritical core supercritical and in a consistent way with our simplification.

Lastly, we take a look at the magnetic diffusivities in the outflow cavity, which is the focus of Paper II. Observations of protostellar jets have found evidence of ionized material \citep[see e.g.,][]{Moscadelli2021multi, Rodriguezkamenetzky2017, Carrasco-Gonzalez2021, Guzman2010}, possibly from shock ionization. As discussed in \cite{Marchand2016,Marchand2022}, ionization leads to a general decrease in the magnetic diffusivities. So far, the models of ambipolar diffusion and the Hall effect done in e.g., \cite{Marchand2016,Marchand2022} and \cite{Tsukamoto2020} do not consider the shock ionization in the outflow cavity. Material in the outflow cavity and close to the massive protostar is very low density and is in presence of strong magnetic fields, which means that neglecting shock ionization would overestimate magnetic diffusivities coming from the ambipolar diffusion and Hall effect terms, when comparing to the curves by \cite{Marchand2022}. By only considering Ohmic dissipation (for which the resistivity is low for low densities and independent of magnetic field strength), the material in the cavity behaves closer to ideal MHD theory, in agreement to what has been found from observational evidence (see Sect. \ref{S: obs} and \cite{Moscadelli2022}). This can be seen in Fig. \ref{brho}, where the magnetic field strengths in the cavity (light-colored dots) are able to increase while keeping the values constrained in the disk (dark-colored dots). A realistic treatment of the problem, however, would require the consideration of all forms of magnetic diffusivity (which would probably impact the disk dynamics) and the effect that thermal ionization and shock ionization have on the magnetic diffusivities corresponding to the outflow cavity.

\subsection{Density and temperature profiles}
\begin{figure}
	\includegraphics[width=\columnwidth]{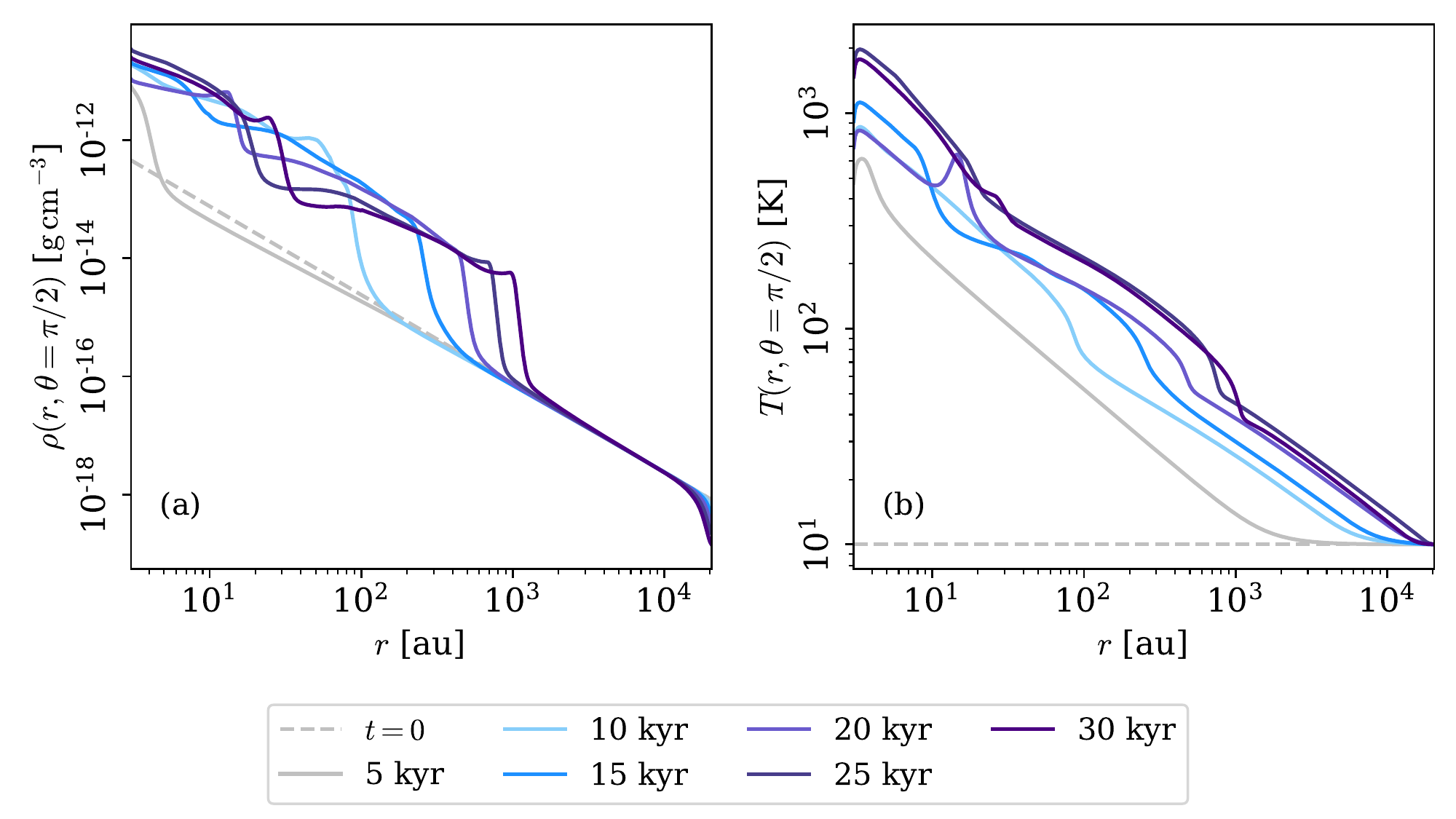}
	\caption{(a) Density and (b) temperature profiles on the midplane, for the fiducial simulation on grid x8.}
	\label{thindisk_denstemp}
\end{figure}

\begin{figure}
	\centering
	\includegraphics[width=\columnwidth]{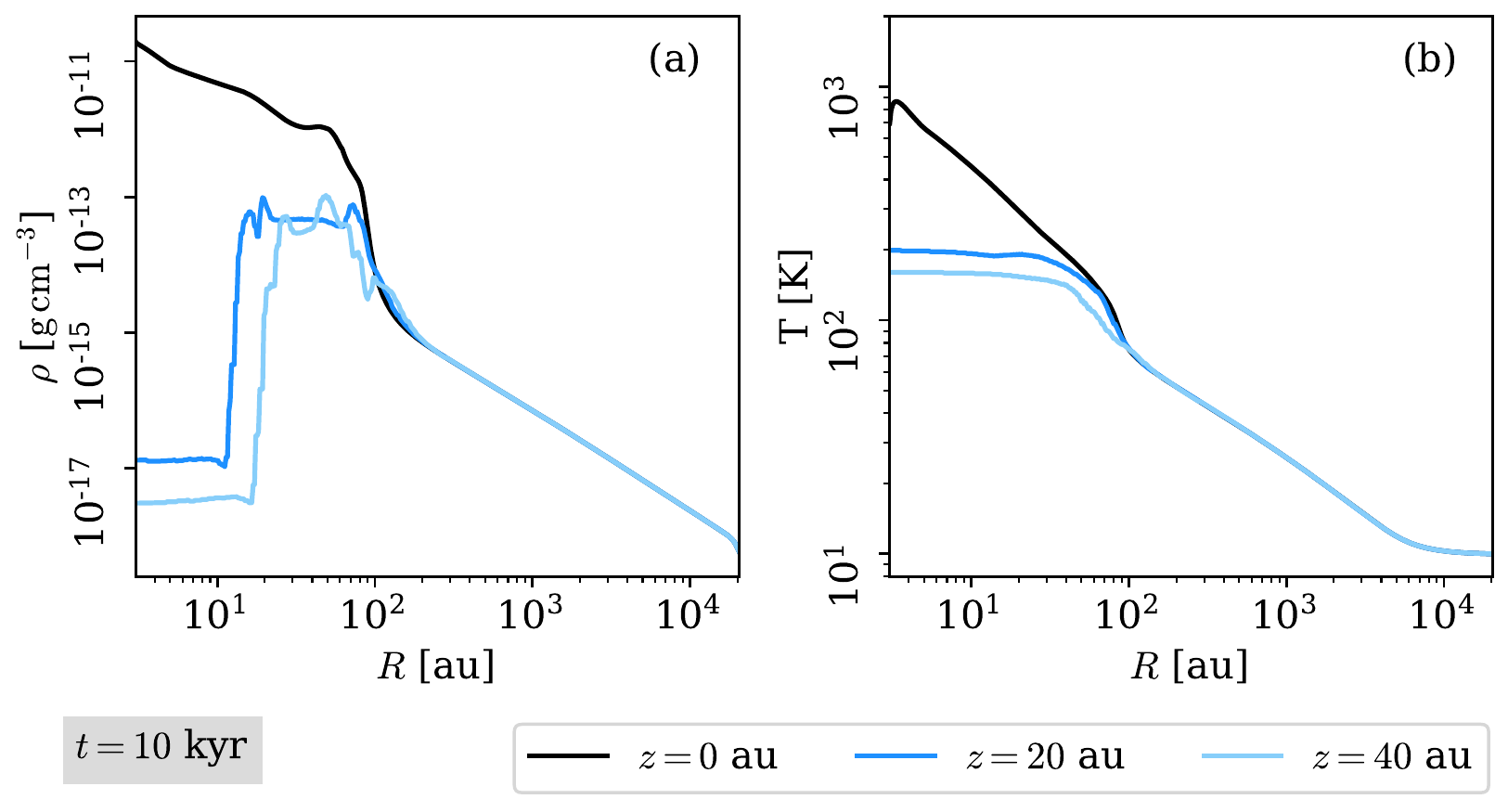}
	\caption{(a) Density and (b) temperature for two different altitudes in the thick layers of the disk (blue lines), and the corresponding profile for the thin layer (black). The data corresponds to the fiducial simulation on grid x8.}
	\label{thickdisk_denstemp}
\end{figure}

Figure \ref{thindisk_denstemp} presents the density and temperature profiles for the midplane. The thin layer has densities over $10^{-16} \unit{g\,cm^{-3}}$, and they reach over $10^{-11}\unit{g\,cm^{-3}}$ at the inner boundary. For $t\lesssim 15\unit{kyr}$, the density profile of the thin layer follows an approximate power law. Later, the magnetically-braked region is formed, corresponding to the high density feature observed for $r \lesssim 30 \unit{au}$ in Fig. \ref{thindisk_denstemp}a. Some of the material of the magnetically-braked region is drawn from the thin layer of the disk: the density profile of the thin layer reveals the existence of a plateau close to its magnetic braking radius, which coincides as well with the decrease in the azimuthal contribution to the specific kinetic energy due to magnetic braking (Fig. \ref{thindisk_energies}c). However, another part of the material in the magnetically-braked region corresponds to the inflows from the cavity wall and thick layer of the disk, both of which also experience magnetic braking and continuously replenish it.

The temperature profile of the thin layer follows roughly a power law of the form $T\propto r^{-1/2}$, and it experiences a general increase over time. The transition between the disk and the envelope is accompanied by a change in the temperature gradient; the transition between the magnetically-braked region and the thin layer of the disk is more subtle. We remind the reader that the present simulations do not include the effect of irradiation from the star, for which we anticipate two additional features not present in Fig. \ref{thindisk_denstemp}b, namely, an increase in temperature due to the flux from the forming massive star, and the existence of the dust evaporation front \citep[see][]{Kuiper2010circ}. Those features are only relevant for $t\gtrsim 20 \unit{kyr}$.

Fig. \ref{thickdisk_denstemp}a presents the density profile at two different altitudes crossing the thick layer of the disk at the time $t=10\unit{kyr}$ (when $M_\star = 3 \unit{M_\odot}$), 
 accompanied by a comparison with the density in the midplane. Differently than the thin layer of the disk, the thick layer has a nearly uniform density of $\sim 10^{-14}\unit{g\, cm^{-3}}$. The density drop for $r\lesssim 10 \unit{au}$ corresponds to the outflow cavity. The mean density in the thick layer decreases slightly with time, however, it roughly remains within the same order of magnitude throughout the simulation.

Because the density is nearly uniform, the temperature of the thick layer of the disk remains also relatively uniform, at around $200\unit{K}$ in comparison to the power-law temperature of the thin layer. This means that it should be possible in principle to observe the vertical stratification of the disk by distinguishing the line emission from each layer.

\section{Dependence of the resulting disk properties on initial cloud properties} \label{s:parameter-study}

\begin{figure*}
	\centering
	\includegraphics[width=\textwidth]{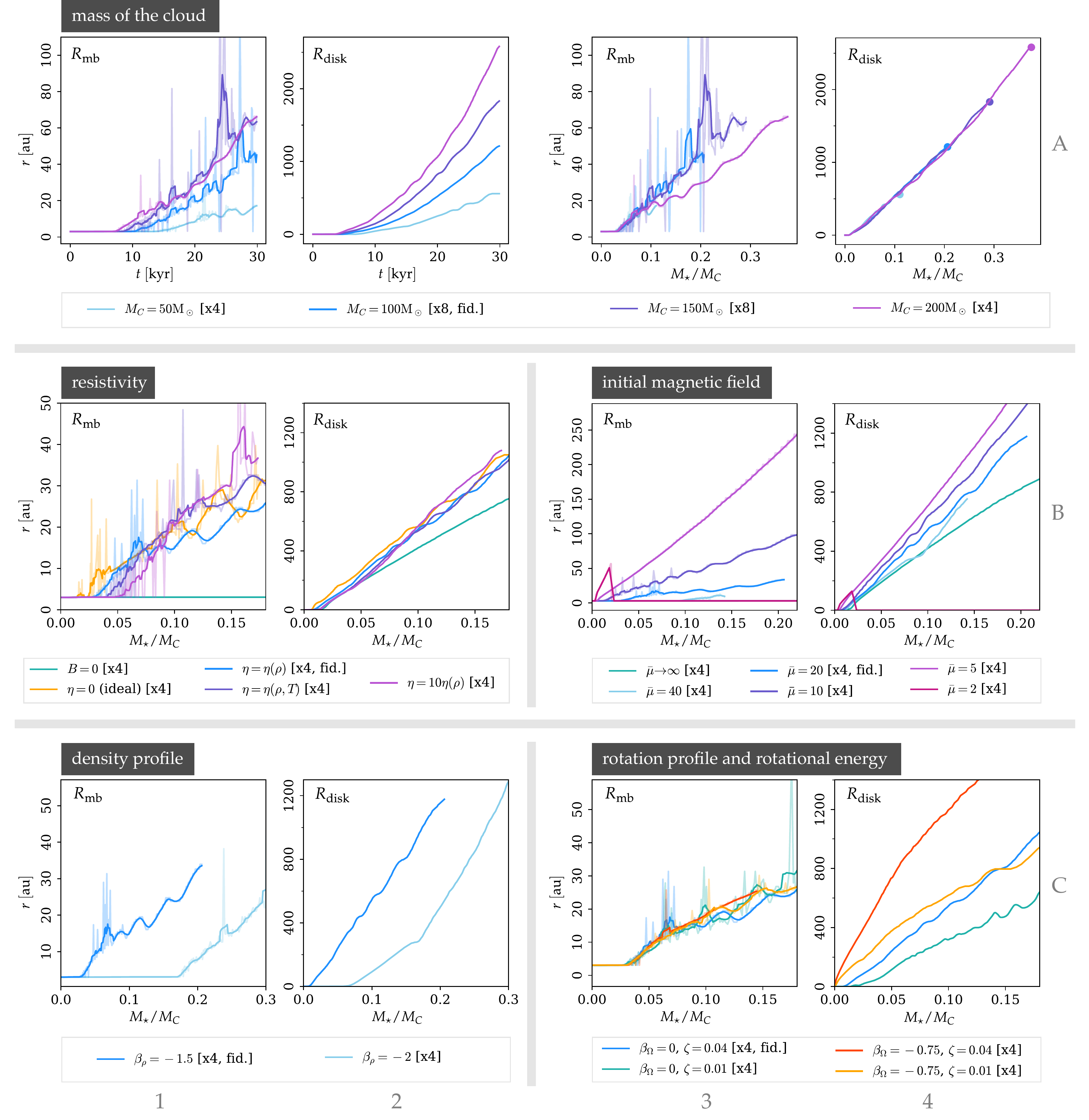}
	\caption{Radius of the disk ($R_\text{disk}$) and its magnetic braking radius ($R_\text{mb}$) for different initial values of the mass of the cloud, magnetic field, density profile, rotational profile, rotational energy and resistivity models. The transparent lines in the panels that show $R_\text{mb}$ indicate the full data, while the solid lines show the moving average. The colored dots in panel A4 indicate $t=30\unit{kyr}$ in each simulation. For all the panels in rows B and C, $M_C=100\unit{M_\odot}$ was used.}
	\label{rdisk}
\end{figure*}

\begin{figure}
	\centering
	\includegraphics[width=\columnwidth]{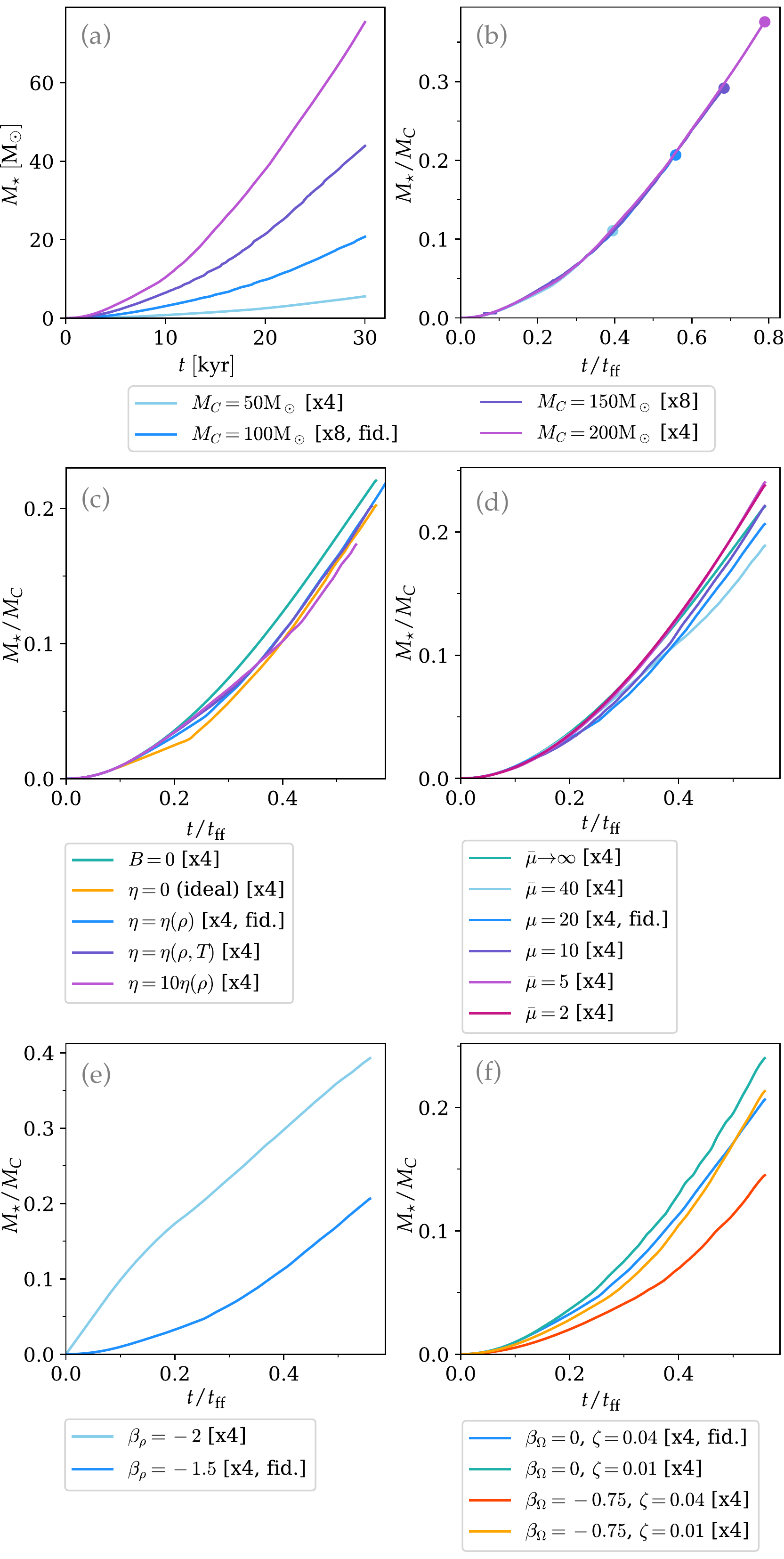}
	\caption{Mass of the protostar, corresponding to the mass in the sink cell, for several values of the (\emph{a}--\emph{b}) mass of the cloud, (\emph{c}) resistivity model, (\emph{d}) mass-to-flux ratio, (\emph{e}) density profile, and (\emph{f}) rotation profile and rotational energy. Panel \emph{b} shows the same results as panel \emph{a}, but normalized in such a way that scalability can be readily seen. For panels (c)--(f), $M_C = 100\unit{M_\odot}$ and $t_\text{ff} = 53.73\unit{kyr}$.}
	\label{mstar}
\end{figure}

\begin{figure}
	\centering
	\includegraphics[width=\columnwidth]{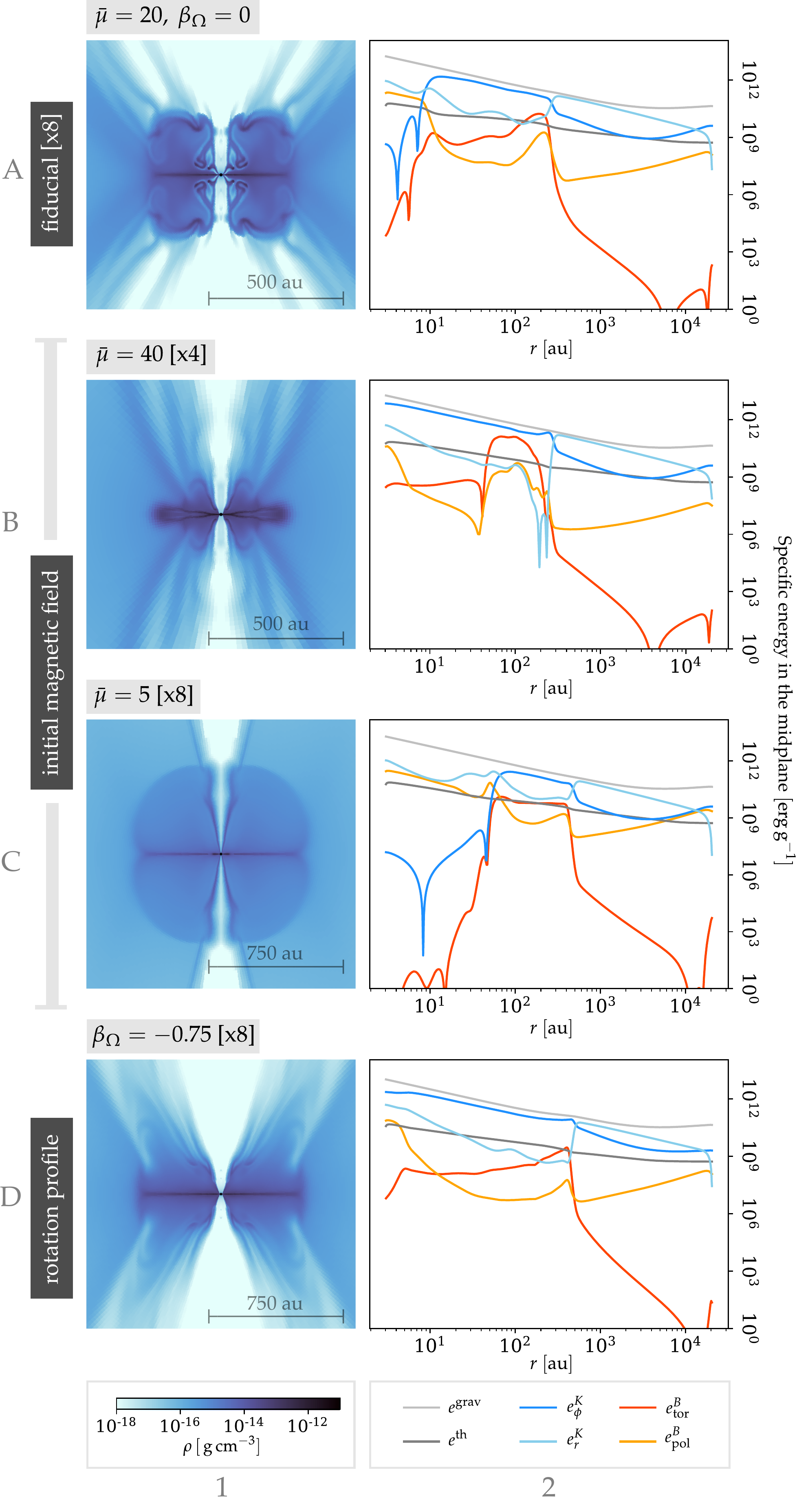}
	\caption{Density structure (column 1) and specific energies (column 2) of the disk for selected values of the normalized values of the mass-to-flux ratio (rows B and C), and rotation profile (column D). All the snapshots were taken at $t=15\unit{kyr}$ of evolution. All cases consider a rotational-to-gravitational energy of 4\%.}
	\label{disk_morph}
\end{figure}

We perform a series of simulations to explore the large parameter space of initial conditions. We vary the fiducial model model described above in terms of the initial mass of the cloud core, its density and rotation profiles, its rotational-to-gravitational energy ratio, its initial magnetic field strength and probe the effect of different models of Ohmic resistivity. In this section, we discuss how those changes affect the processes discussed in Sect. \ref{S: dynamics}. As a probe, we use the radial extent of the Keplerian-like disk (between the magnetic braking radius and the [outer] radius), defined as the region where the material is within $\pm 25\%$ Keplerianity, and the results of which are presented in Fig. \ref{rdisk} for the whole parameter space. Similarly, the mass of the forming massive star is presented in Fig. \ref{mstar} for the whole parameter space. We also examine differences in the density structure of the disk for selected values of the parameters, and relate them to the dynamical information revealed by the specific energy; the results are shown in Fig. \ref{disk_morph}.

\subsection{Dependence on cloud core mass} \label{S: disk var M}

Cloud cores of masses of $50, 100, 150$ and $200 \unit{M_\odot}$ were considered, which form stars ranging masses from $6$ to $75\unit{M_\odot}$ after $30\unit{kyr}$ of evolution (see Fig. \ref{mstar}a). Further evolution and additional physical effects which are not considered here would be necessary to determine the final mass of the star. When changing the initial core mass of the fiducial model, we also change the normalization of the rotation profile in order to keep the rotational-to-gravitational energy ratio as in the fiducial model. The mass and size of the cloud core determines its free-fall timescale
\begin{equation}
	t_\text{ff} = \left(\frac{\pi^2 R_C^3}{8GM_C}\right)^{1/2},
\end{equation}
which is $52.4\unit{kyr}$ for the fiducial case of $M_C = 100\unit{M_\odot}$ and $R_C = 0.1\unit{pc}$. Panels A1 and A2 of Fig. \ref{rdisk} show that the disk and the magnetically-braked region are smaller for the lower-mass cases when plotted against time, however, when investigated as a function of the stellar mass as a fraction of the initial mass of the cloud (panels A3 and A4), both the size of the disk and the magnetic braking radius coincide well. We find that resulting curves roughly follow the empirical linear relation
\begin{equation}
 R_\text{disk}(m) \approx R_\text{onset} + R_\infty \left( m - m_\text{onset} \right) ,
\end{equation}
where we define
\begin{equation}
m \equiv \frac{M_\star}{M_C}.
\end{equation}
$R_\text{onset} \equiv 3\unit{au}$ is the minimum radius of the disk we can detect in the simulation data (the size of the sink cell), $m_\text{onset}=0.016$ is the corresponding normalized mass of the protostar for $R_\text{onset}$, and $R_\infty=6380\unit{au}$ can be interpreted as the approximate value of the radius of the disk at the hypothetical end of the collapse where $m\sim 1$. This latter value will never be reached in reality because of stellar feedback: \cite{KuiperHosokawa2018}, e.g., found that the disk lifetime is expected to be of only a few $10^5\unit{yr}$ (see also  \citealt{Kee2019} for the conditions for limiting accretion through the disk). A similar empirical disk size formula was previously offered in \cite{Kuiper2011}.

We also verified that the mass of the protostar, expressed as a fraction of the mass of the cloud, scales with time presented as fraction of the free-fall timescale (cf. panels \emph{a} and \emph{b} of Fig. \ref{mstar}). This result indicates that despite the differences in the scale-dependent microphysics, namely, the radiative transfer, resistivity and self-gravity, the disk scales well with the free-fall timescale, which allows our results to be used with a more ample range of cloud masses than what is presented here. Defining $\tau \equiv t/t_\text{ff}$, we also find an approximate fit for the mass of the protostar with a power law:
\begin{equation}
  m(\tau) \approx k_1\, \tau^{k_2},
\end{equation}
where $k_1 \approx 0.56$ and $k_2 \approx 1.8$.

\subsection{Ohmic resistivity} \label{S: disk var eta}

The Ohmic resistivity model controls the diffusivity of the plasma, and therefore it has an impact on magnetic braking. We studied the effects of using the resistivity model of \cite{Machida2007} in two versions: the full expression dependent on both density and temperature, and the same expression but with a fixed temperature of $º   T=10\unit{K}$, as it was used in the isothermal study of \cite{Anders2018}. We denote the first one as $\eta(\rho,T)$ and the second one as $\eta (\rho)$ given the near linear behavior of the resistivity in the range of densities for which we are interested. We estimate that the version that considers the temperature dependence differs at most by a factor of $100$ in regions close to the protostar with respect to the version with the fixed temperature. We compare our results as well with the ideal MHD case ($\eta = 0$) and the non-magnetic (hydrodynamical) case ($\eta\to \infty$). In order to mimic the magnetic diffusivity expected in the disk region if ambipolar diffusion were taking into account, we also include a simulation with an artificially high Ohmic resistivity (ten times the temperature-independent value, denoted as $10\eta(\rho)$). This factor is motivated by the difference in the curve for ambipolar diffusivity and Ohmic resistivity in \cite{Marchand2022}, for disk densities. There is an additional difference between the simulations of this series: while most simulations of this study consider a constant dust and gas opacity of $1\unit{cm^2\, g^{-1}}$, the simulation with $\eta(\rho,T)$ uses the dust opacity table from \cite{LaorDraine1993} and a constant gas opacity of $10^{-2}\unit{cm^2\, g^{-1}}$.

For the rest of Sect. \ref{s:parameter-study}, we present the size of the disk as a function of the stellar mass as a fraction of the cloud mass, instead of time, because the calculation of the offset from Keplerianity (which defines the disk) is dependent on the gravity of the (proto)star, and by fixing it we are able to compare and assess the effects of the other terms that influence the radial equilibrium of the disk, namely, the angular momentum and pressure support. Additionally, our results can be scaled as a result of the discussion in Sect. \ref{S: disk var M}.

In the panel B1 of Fig. \ref{rdisk}, we observe that the magnetically-braked region develops earlier (when the mass of the star is lower, cf. Fig. \ref{mstar}c) in the ideal MHD case, as compared to the corresponding curves for the simulations that consider resistivity. The case with $\eta(\rho)$ requires a more massive protostar (more time evolution) for the magnetic braking radius to appear, a trend that is confirmed with the simulations that consider higher resistivity: the temperature-dependent resistivity delays the appearance of magnetic braking, while the simulation with the artificial factor of ten in resistivity delays it even further. As expected, the disk in the case with no magnetic fields does not develop a magnetic braking radius. A disk is able to form at all in the ideal MHD case because the relatively weak magnetic field considered for this parameter scan.

From this evidence, we infer that the presence of resistivity delays the action of magnetic braking compared to the results only considering ideal MHD. Higher resistivities delay magnetic braking more, however, as discussed in Sect. \ref{S: disk MB}, because magnetic diffusion does not completely suppress the toroidal magnetic field, the Lorentz force continues to grow until enough magnetic tension force is built to decelerate the material. As a result, the innermost parts of the disk lose their centrifugal support.

The disk radius (Fig. \ref{rdisk}B2) is in general larger in the magnetic cases in comparison to the non-magnetic case. Moreover, the disk radius of the run with no resistivity (ideal MHD) is marginally larger than the ones with resistivity. Magnetic braking removes angular momentum from the disk because of the torque produced by magnetic tension; therefore, a smaller disk would be expectable. However, magnetic tension is highest in the inner disk, where the magnetic field lines are wound the most, and not in the outer disk. As discussed in Sect. \ref{S: dynamics - PB}, magnetic pressure can provide an additional radial support for the disk. This means that in the ideal MHD case, where the toroidal magnetic field is high in the disk, the additional magnetic pressure supports a larger disk. Magnetic diffusion lowers the toroidal magnetic field, and therefore the disk does not have radial magnetic pressure support: this can be seen at early times (where $M_\star < 5\unit{M_\odot}$ or $m = 0.05$) where the simulations with $\eta(\rho,T)$ and $10\eta(\rho)$ yield a disk radius that almost matches the one obtained in the non-magnetic case. As the toroidal magnetic field increases over time due to the winding of magnetic field lines, the magnetic pressure support also increases and the disk becomes larger than the non-magnetic case.

Given that magnetic braking delivers mass onto the protostar, we investigate the role of resistivity in the mass of the forming massive star. The corresponding curves for the mass of the protostar as a function of time are shown in Fig. \ref{mstar}c. The results for the resistive and non-resistive cases are very similar. The stellar mass, however, increases more in the non-magnetic case in comparison to the magnetic cases. This happens because the protostar in the non-magnetic case accretes material not only through the disk, but also through the infalling envelope along the bipolar directions. On the contrary, in the magnetic cases, there are magnetic outflows that impede accretion through the axis, resulting in a slightly lower mass.

\subsection{Initial magnetic field strength} \label{S: disk var B}

We study values of the normalized mass-to-flux ratio ranging from $\bar \mu = 2$ (strongest magnetic field) to $\bar\mu = 40$ (weakest magnetic field).%

The results of the radial extent of the centrifugally-supported disk are shown in the panels B1 and B2 of Fig. \ref{rdisk}. The curve for $\bar \mu = 2$ goes quickly to zero, which means that the structure formed is strongly sub-Keplerian and it is not classified as a Keplerian-like disk according to our criterion. A more detailed study of the specific energies revealed that the structure has a comparable radial and azimuthal velocities. Our result of no disk for $\bar \mu = 2$ should be taken with caution, because we do not include ambipolar diffusion in our calculations. As ambipolar diffusivities increase with magnetic field (contrary to Ohmic dissipation), the cases with strong magnetic fields are missing magnetic diffusivity that may allow the disk to form (see the discussions in Sects. \ref{S: other magn diff} and \ref{S: previous amb diff}).

The other values of the mass-to-flux ratio confirm that magnetic braking is responsible for the formation of the magnetically-braked region because higher initial magnetic fields build enough magnetic tension earlier. Additionally, the magnetically-braked region is larger for stronger magnetic fields. The radius of the disk, on the other hand, is larger for stronger magnetic fields, which supports the hypothesis that higher magnetic pressure provides additional support against gravity and allows material from larger radii (with lower angular momentum) to be part of the disk. As for the accretion onto the protostar, Fig. \ref{mstar}d reveals that higher magnetic fields do tend to deliver more mass onto the protostar, due to stronger magnetic braking at late stages.

The rows A--C of Fig. \ref{disk_morph} present a comparison of the morphological differences of the disk when it has evolved for $t=15\unit{kyr}$ under different values of the mass-to-flux ratio. For $\bar \mu = 40$, the weaker magnetic field produces a disk with a thick layer that is smaller than the thin layer. Both regions of the disk are of the same size for $\bar \mu = 20$, which confirms magnetic pressure as the nature of the vertical support in the thick disk. However, the part of the thin layer that is not enveloped by the thick layer of the disk becomes inflated by magnetic pressure, as the curve for the toroidal contribution to the specific magnetic energy reveals in Fig. \ref{disk_morph}B2: in the outer disk ($r\sim 100\unit{au}$), $e^B_\text{tor}$ becomes higher than $e^\text{th}$. Nevertheless, a similar analysis of the specific energies at later times reveals that the thick layer of the disk grows over time until it completely envelops the thin layer (which happens at around $t\sim 19\unit{kyr}$). At $t\approx 25.5\unit{kyr}$, the thin and thick disk layers of the disk of the case with $\bar \mu = 40$ become morphologically and dynamically similar to the results shown in panels A1 and A2 corresponding to the $\bar \mu = 20$ at $15\unit{kyr}$; the difference being the size of the disk and the mass of the protostar. For a stronger initial magnetic field ($\bar \mu = 5$, row C of Fig. \ref{disk_morph}), the thick layer becomes thicker because of the increased magnetic pressure.

\subsection{Initial density profile}
A steeper initial density profile accelerates the accretion onto the protostar at early times because of the increased concentration of mass (and therefore gravity) in the center of the cloud. Apart from a density profile with $\beta_\rho = -1.5$, we also tried a profile with $\beta_\rho = -2$. The mass of the protostar (Fig. \ref{mstar}e) increases rapidly at early times with $\beta_\rho = -2$, but the accretion rate decelerates over time, which is the opposite behavior observed for $\beta_\rho = -1.5$. This behavior can be readily seen from the power-law fits to those curves: the normalization constant is similar in both cases, but while the fiducial case is described by $m \propto \tau^{1.80}$, the steeper initial density profile exhibits $m \propto \tau^{0.94}$. The  radius of the disk (Fig. \ref{rdisk}C2) is larger in the case of the shallower density profile when seen as a function of the stellar mass and its behavior is longer linear. However, given that the stellar mass increases more rapidly in the case of the steeper density profile, in this latter case, a disk of a given size is observed earlier in time. A similar comparison of stellar masses and time reveals that magnetic braking happens at roughly the same time for both simulations, because the magnetic field is wound by rotation in a similar way in both cases. While the magnetically-braked region appears at $M \approx 3\unit{M_\odot}$ when $\beta_\rho = -1.5$, it appears at $M\approx 17\unit{M_\odot}$ for $\beta_\rho = -2$; however, the protostar reaches both masses at the same time, $t \approx 0.2 t_\text{ff} \approx 10.7\unit{kyr}$. As a result, we conclude that a disk that does not exhibit a magnetically-braked region at a given time can be obtained for a range of protostellar masses by adjusting the initial density profile.

\subsection{Initial rotation profile} \label{S: disk var rotprof}

We investigated the effect of two scenarios: initial solid-body rotation and a rotation profile that scales with the cylindrical radius at $\Omega \propto R^{-\beta_\Omega}$, with $\beta_\Omega = \beta_\rho/2 = -0.75$. This choice of $\beta_\Omega$ keeps the initial ratio of rotational to gravitational energy independent of the radius of the cloud. Additionally, the particular value of $-0.75$ produces a cloud for which the offset from Keplerianity is uniformly negative. We complemented this parameter scan with an investigation of the effects of using a lower initial rotational-to-gravitational energy ratio $\zeta$ of 1\% instead of 4\% as in the fiducial case.

An inspection between panels D1 and A1 of Fig. \ref{disk_morph} reveals that the thick layer of the disk is flatter in the cloud with the steep rotation profile. The higher angular velocity at the center of the cloud allows for the formation of a larger accretion disk earlier in the simulation, in comparison with the cloud initially rotating as a solid-body. Panel C4 of Fig. \ref{rdisk} confirms this: the curves for the disk radius for $\beta_\Omega = -0.75,\,\zeta = \{0.01,0.04\}$ are in general higher than the corresponding curves for $\beta_\Omega = 0$, and resemble a power law rather than a linear function as it is the case for solid-body rotation. For a fixed value of $\beta_\Omega$, the disk becomes larger with more content of initial rotational energy. Additionally, both curves for $\beta_\Omega = -0.75$ show that the disk forms at very early times during the simulation compared to $\beta_\Omega = 0$: for the solid-body case, angular momentum conservation during the gravitational collapse provides the necessary increase of angular momentum in the center of the cloud in order to build up a disk, while the steep rotation profile already provides some of this angular momentum from the start. Interestingly, the cases $\beta_\Omega = -0.75,\,\zeta = 0.01$ and $\beta_\Omega = 0,\,\zeta = 0.04$ produce a similar disk radius. The mass of the formed protostar (Fig. \ref{mstar}f) is lower in the cases with a steep rotation profile and higher initial rotational energy. This is consequence of the earlier formation of a disk that is also larger (and therefore more massive), at it reduces the efficiency of accretion onto the protostar (in \citealt{Oliva2020}, we utilized this fact to study disk fragmentation in the absence of magnetic fields). The magnetic braking radius of the disk is very similar for all cases when plotted as a function of the protostellar mass, however, given that the masses grow different in time, magnetic braking appears later in the case with $\beta_\Omega = 0,\,\zeta = 0.04$.

From the dynamical point of view, a comparison between panels D2 and A2 of Fig. \ref{disk_morph} reveals that in general, the different contributions to the specific energy have a more uniform radial gradient. Given that the offset from Keplerianity in the midplane can also be computed as $1-2e^{K}_\phi/e^\text{grav}$, the disk formed with the steep rotation profile is more uniformly Keplerian. This is also manifested when comparing the curves for $e^{K}_r$ inside of the disk in both cases.

\section{Comparison to observations} \label{S: obs}

\subsection{Natal environment of massive star formation}
The parameter space in our simulation set was motivated by observational typical values found in regions of massive star formation. Surveys of cloud cores in the early stages of massive star formation have found power-law density profiles with exponents ranging in $1.5\lesssim -\beta_\rho \lesssim 2.6$ \citep[see, for example,][]{2022AA...657A...3G,Beuther2018, 2002ApJ...566..945B, 2002ApJS..143..469M, 2003A&A...409..589H}, with fragmenting cores having typical values of around $\beta_\rho \approx -1.5$. We take values in the extremes of this interval.

For the initial angular momentum, we start from the results of the survey by \citep{Goodman1993}, who found that the the ratio of rotational to gravitational energy of cloud cores ($\zeta$) is of a few per cent, with typical values of around 2\%. Those authors fitted linear gradients to the cloud cores that had signs of rotation, and found solid-body profiles. Our second choice for the initial rotation profile, $\beta_\Omega = \beta_\rho/2$, is motivated by keeping $\zeta$ independent of the radius of the cloud core, as explained in Sect. \ref{S: disk var rotprof}.

The values of the mass-to-flux ratio are motivated by the supercriticality in star-forming cloud cores \citep[e.g.][]{Beuther2020, Girart2013} and the following analysis. In low-mass prestellar cores, the magnitudes of the magnetic field have been found to be of the order of a few to tens of micro Gauss \citep{2022Natur.601...49C, 2008ApJ...680..457T}. As the pre-collapse conditions in the massive star formation case are currently not known, we consider two scenarios (see also the discussion in \citealt{Machida2020}):
\begin{enumerate}
\item That the normalized mass-to-flux ratio is independent of mass, which means that magnetic field strengths of 100 to $1000\,\mu\mathrm{G}$ are required in the cloud prior to the gravitational collapse.
\item That the magnetic field strength of the high-mass star formation case is similar to the low-mass counterpart, which implies that the mass-to-flux ratio must be high.
\end{enumerate}
The first case is considered in the low mass-to-flux ratios in our parameter space ($\bar \mu \sim 5$), while the second case corresponds to high mass-to-flux ratios, including our fiducial case ($\bar \mu = 20$).

\subsection{Size of the disk}
The analysis in Sect. \ref{s:parameter-study} has shown that the radius of the disk is more strongly determined by the initial density and rotation (angular momentum) profiles of the large scale initial condition, as compared to the magnetic field and resistivity models, within the ranges of those values expected from observations. Panel B4 of Fig. \ref{rdisk} shows that when $M_\star/M_C \sim 0.15$, the average $R_\text{disk}$ for the values of $\bar \mu$ we considered is around $850\unit{au}$, with a variation of the order of $\pm 200\unit{au}$. However, the radius of the disk and its growth rate vary much more with the density profile and the rotation profile (panels C2 and C4 of Fig. \ref{rdisk}). Smaller disks are produced with a lower and flatter initial distribution of angular momentum, and with shallower initial density profiles. This conclusion is crucial for the comparison of our results with previous studies, which is done in Sect. \ref{S: previous}.

With regards to the available observational evidence of disk-jet systems around forming massive stars, we see that in principle it is possible to obtain a variety of disk sizes that are not only dependent on the age of the system but also on the initial distribution of matter and angular momentum, and not so strongly determined by magnetic braking in the outer regions of the disk. For example, assuming that the molecular cloud has the typical values of $M_C = 100\unit{M_\odot}$ in a sphere of radius $0.1\unit{pc}$ of our fiducial case, we obtain a disk of radius $\sim 1200 \unit{au}$ when the mass of the protostar is $M_\star \sim 10 \unit{M_\odot}$, if we assume a steep and fast initial rotation profile ($\beta_\Omega = -0.75$, $\zeta = 0.04$), which would provide a possible initial configuration for the observations of HH 80-81 \citep{Girart2017}.

In \cite{Moscadelli2022}, we utilized the fiducial simulation run on grid x16 to model the accretion disk and jet of IRAS 21078+5211. For that particular system, observations of molecular rotational transitions had revealed the existence of a Keplerian-like accretion disk of around $200 \unit{au}$ in size \citep{Moscadelli2021multi}. In order to find a model of our catalog that was consistent with the observations, in \cite{Moscadelli2022} we used the mass of the protostar, the radius of the disk, and the morphology of the ejected material to constrain the parameter space and the time elapsed. In that study, we found strong agreement between the velocity field of the observed water masers and the velocity streamlines predicted by the chosen model from our catalog. This good agreement in turn constitutes evidence that the results for the size of the disk as a function of the pre-collapse conditions --which we present in Sect. \ref{s:parameter-study}-- are in line with what is expected from observations. We note, however, that the wide streamline traced by the maser points (Fig. 2 of \citealt{Moscadelli2022}) is slightly closer to the assumed disk plane compared to the reference wide streamline from the simulations. This can be caused by a number of factors, for example, a slightly different assumed perspective between the simulations and the observations, but a possible explanation is that the disk in IRAS 21078+5211 is thinner than what is predicted in the simulations.

\section{Comparison with previous numerical studies} \label{S: previous}

The present study is a continuation of the work started by \cite{Anders2018}. Their simulations correspond to our fiducial case but with a grid equivalent to our x2 grid, and used an isothermal equation of state. An isothermal equation of state yields a much smaller pressure scale height of the disk or thinner disk, respectively. In combination with the rather coarse grid resolution, the thickness of the disk could not be resolved on the numerical grid, making an in-depth convergence study problematic. In that sense, the addition of radiation transport and higher resolutions has enabled us to resolve the thin layer of the disk, and clearly differentiate it from the surrounding region that we call the thick layer of the disk. Also, they do not report about the existence of the magnetically-braked region, however, it is visible upon close examination of their Fig. 21 as a region where the vertical velocity is negative close to the center of the disk. This was originally thought to be a numerical effect caused by the additional mass created by the Alfvén limiter, however, we performed a parameter scan with several values of the Alfvén limiter, the highest values of which produce negligible artificial mass, and found a magnetically-braked region in all of them.

\subsection{Studies with ideal MHD}
\cite{Banerjee2007}, \cite{Seifried2011}, \cite{Myers2013} and \cite{Rosen2020} conducted simulations of the formation of massive stars under the assumption of ideal MHD. Although \cite{Banerjee2007} observed magnetically-driven outflows, their disk is always sub-Keplerian. \cite{Myers2013} started from a cloud of $300 M_\odot$, an initial velocity profile with supersonic turbulence and no rotation, and $\bar \mu = 2$ with $B_z \propto R^{-1/2}$. They included grey radiative transfer and radiative protostellar feedback, and used a 3D AMR grid with a maximum resolution of $10 \unit{au}$, as well as an additional isothermal high resolution run ($\Delta x = 1.25 \unit{au}$). They only find a disk in their high resolution run; it has a radius of $40\unit{au}$ when $t\sim 0.2 t_\mathrm{ff}$, for which $M_\star \sim 3.5 M_\odot$. It is not trivial to compare their results to ours because of the difference in initial velocity fields, and the use of an isothermal equation of state. Using the scalability of our results with the mass of the cloud, we find the radius of the disk to be $128\unit{au}$ for $t = 0.2 t_\mathrm{ff}$ in the simulation for $\bar \mu = 2$. However, if we take $M_\star/M_C$ and compute the corresponding fraction of the free-fall timescale, we get $t/t_\mathrm{ff} \approx 0.104$, for which our disk has a radius of $\sim 30 \unit{au}$, which is in line with the fact that supersonic turbulence delays the formation of an accretion disk as enough angular momentum has to assemble close to the forming star. \cite{Rosen2020} also considered a cloud core with supersonic turbulence and radiation transport, however their coarse grid (minimum cell size of $20\unit{au}$) only allows them to partially resolve a disk-like structure and therefore a direct comparison to our results is not possible.

\cite{Seifried2011} considered a setup similar to ours in terms of the mass and radius of the cloud, as well as its initial density and rotation profiles, but used a cooling function instead of radiation transport, and only considered the ideal MHD approximation. The code FLASH with an AMR grid of maximum resolution of $4.7\unit{au}$ was used. The authors explored several values of $\bar \mu$ ranging from 2.6 to 26, but with a uniform plasma $\beta$, and obtained a Keplerian-like accretion disk for $\bar \mu = 26$. They observe a drop in Keplerianity for the inner parts of the disk in the case of $\bar \mu = 10.4$, and credit it to magnetic braking, in agreement with our results; however, they do not observe a disk for $\bar \mu < 5.2$ because the high density of the thin layer of the disk decouples the gas from the magnetic field im resistive MHD runs.

\subsection{Studies including Ohmic resistivity}
\cite{Matsushita2017} and \cite{Machida2020} use the same Ohmic resistivity model as we do \citep{Machida2007}, but both consider a barotropic equation of state instead of solving for radiation transport. In \cite{Matsushita2017}, a disk was formed, however, their analyses are strongly focused on the magnetic outflows. \cite{Machida2020} consider cloud cores of radius $0.2\unit{pc}$ with masses ranging from $11$ to $545$ solar masses, an enhanced Bonnor-Ebert density profile, and slow initial solid-body rotation. If we expand our cloud to $0.2 \unit{pc}$, keeping the same density profile as in the fiducial case, the mass of the cloud becomes $283 \unit{M_\odot}$, which means that our setup is similar to their simulation series E. The authors also make a parameter scan with the  normalized mass-to-flux ratio with values ranging from 2 to 20, and use a three-dimensional AMR grid with a maximum resolution of $0.62\unit{au}$. After $10\unit{kyr}$ of evolution, they find disk radii of a few hundred astronomical units, which develops spiral arms and fragments for some configurations.

\subsection{Studies including ambipolar diffusion} \label{S: previous amb diff}

\cite{Mignon-Risse2021} and \cite{Commercon2022} studied the formation of massive stars from $100 \unit{M_\odot}$ cloud cores of radius $0.2\unit{pc}$, with a centrally-condensed initial density profile; which roughly correspond to our configuration $M_C = 50\unit{M_\odot}$. The cloud cores are initially in solid body rotation and are threaded by a uniform magnetic field determined from the normalized mass-to-flux ratio $\bar \mu = 5 \text{ and } 2$. They treated radiation transport with the flux-limited diffusion approach and gray stellar irradiation; for magnetic diffusivity, they considered ambipolar diffusion using the model from \cite{Marchand2016} but no Ohmic resistivity, and ran their simulations on a three-dimensional AMR grid with the code RAMSES down to a maximum resolution of $5\unit{au}$. In a nutshell, those studies constitute the ones which include the most similar physical ingredients but to our setup but with ambipolar diffusion instead of Ohmic dissipation as the resistive effect and utilizing a different grid method.

Both studies report on thin accretion disks which are vertically supported by thermal pressure, in agreement with our results; however, they do not observe a surrounding thick layer of the disk supported by magnetic pressure. %
Opposite to our findings, the authors find smaller disks with stronger magnetic fields, however, they do find a bigger disk for their ideal MHD run compared to the runs with ambipolar diffusion. In that case, the disk is not only bigger, but also inflated and therefore less dense, in agreement with our results. They attribute the finding of smaller disks to higher ionization in the outer disk, and therefore more magnetic braking. In turn, our results indicate that magnetic braking is highest in the inner disk, where the fluid and the magnetic field lines are dragged faster by rotation, although resistivity causes the dragging to be partial and delayed because of the dominance of magnetic diffusion. Moreover, in our results for the ideal MHD case we observe simultaneously the largest disk and the strongest magnetic braking, because it mostly affects the inner disk.

A comparison of the radius of the disk that we report here and the disk size reported in \cite{Commercon2022} can only be made qualitatively and with several considerations in mind, apart from the difference in the non-ideal MHD effect considered. First, the density profiles in both studies are different because of the presence of an initial density plateau in the inner $\sim 4125\unit{au}$ in \cite{Commercon2022} and \cite{Mignon-Risse2021}, in contrast to our continuous power-law slope down to the smallest scales. This density plateau affects the gravitational collapse and therefore, the elapsed time and mass of the protostar cannot be directly compared. Second, the initial rotational to gravitational energy ratio of the cloud is not the same in the fiducial cases of both studies, but we can roughly compare our $\zeta = 0.04$ case to the ``fast'' case studied in \cite{Commercon2022}, which has $\zeta = 0.05$. This is of special importance because of what is shown in Fig. \ref{rdisk}C4: the radius of the disk is strongly influenced by the initial rotation of the cloud, even more than the initial magnetic field (cf. Fig. \ref{rdisk}B4) or resistivity (Fig. \ref{rdisk}B2). Finally, the disk identification criterion is not the same in both studies: while \cite{Commercon2022} use a set of criteria that involves the radial and vertical dynamics of the disk as well as a density threshold, we consider a purely dynamical criterion (radial equilibrium between gravity and centrifugal force) as it is more usual in observational studies. With this in mind, we note that in our simulation $M_C = 50\, \unit{M_\odot}\ \text{[x4]}$, the  radius of the disk is around $470\unit{au}$ after $27\unit{kyr}$, while their disk is around the same value after $51\unit{kyr}$. However, the mass of the protostar at that time is lower in our case ($4.6\unit{M_\odot}$) than theirs ($9.1\unit{M_\odot}$); both the difference in time and mass can be reasonably attributed to the initial density profile. Additionally, while we find that the disk grows in size with time, most of the nonideal MHD simulations by \cite{Commercon2022} find disks that stop growing at around $100\unit{au}$.

\cite{Commercon2022} and \cite{Mignon-Risse2021} do not report the existence of a magnetically-braked region, for which we offer the following explanations, apart from the different criteria for disk identification we already discussed. First, ambipolar diffusion adds magnetic diffusivity, which may have the effect of delaying the formation of the magnetically-braked region. In more detail: we report on the fact that higher Ohmic resistivities delay the onset of magnetic braking and with it, the formation of the magnetically-braked region. Ambipolar diffusion also decouples the magnetic field from the flow (even though it may act in a different direction than Ohmic resistivity), which leads us to assume that a similar delaying effect on magnetic braking could be obtained upon the inclusion of ambipolar diffusion in our calculations. We note, however, that if this delay in the onset of magnetic braking is longer than the timescale of gravitational collapse of the cloud, the massive star and the disk may form without ever developing a magnetically-braked region. %
Second, spatial resolution: their 3D AMR grid has a minimum cell size (maximum spatial resolution) of $5\unit{au}$, while our grid has minimum cell sizes of the order of $10^{-1}\unit{au}$ close to the sink cell for the simulation series x1, x2 and x4, and $10^{-2}\unit{au}$ for the simulation series x8 and x16. Third, their sink particle algorithm requires the definition of an accretion radius, set to four times the minimum cell size, i.e., $20\unit{au}$, which is the size of the magnetically-braked region for most of the time in our fiducial case. As a consequence, gas orbiting the massive protostar within the accretion radius or in the forming magnetically-braked region cannot be resolved on their finest AMR level while it is resolved on our spherical grid.

Finally, we mention the study by \cite{Masson2016}, who carried out simulations of low-mass star formation considering both ideal MHD and ambipolar diffusion. Although we cannot compare our results quantitatively with theirs, there are several qualitative similarities. The authors observe an accretion disk vertically supported by thermal pressure, with magnetic pressure dominating above the disk. Their figure 14 shows a structure with relatively high angular momentum enveloping the disk for both the diffusive and ideal cases, which is reminiscent of the thick layer of the disk we observe, however, due to the different definitions of the disk used in both studies, we cannot establish a direct correspondence. The same figure also shows that in the ideal MHD case, the inner parts of the disk are destroyed by magnetic braking at late times. Nevertheless, their corresponding run that considers ambipolar diffusion also exhibits for late times a region of low angular momentum (probably due to magnetic braking) in a conical region around the inner disk and that could deliver material to it, in a way that is reminiscent of the early stages of magnetic braking we observe in our simulations. Those results seem to support the idea that increasing the diffusivity leads to a delay but not complete suppression of magnetic braking in a magnetized disk embedded in a collapsing cloud.

\section{Summary and conclusions}

We have modeled the formation of a massive star with a series of 30 magnetohydrodynamical simulations including Ohmic resistivity, radiation transport from thermal emission of the dust and gas, and self gravity. The series of simulations covered a wide range of cloud masses, magnetic field strengths, density profiles, rotation profiles, and ratios of rotational to gravitational energy, in line with currently estimated values from observations. We also performed a convergence study to test the robustness of our results.

After analyzing the fiducial case of our parameter study in depth, we found the following general features of the system:

\begin{itemize}
	\item After the initial gravitational collapse, a Keplerian-like accretion disk is formed.
	\item The accretion disk is divided into two layers: a thin layer supported vertically by thermal pressure, and a surrounding thick layer supported by magnetic pressure. The thin layer of the disk appears only in simulations with sufficiently high resolution.
	\item At early times magnetic diffusion due to Ohmic resistivity is strong in the inner parts of the disk and it greatly reduces magnetic braking there during the magneto-centrifugal epoch ($5 \unit{kyr} \lesssim t \lesssim 15\unit{kyr}$). As time progresses, and the magnetic field lines are continuously dragged by rotation, magnetic braking is observed in the innermost $\sim 50\unit{au}$ of the disk for the fiducial case in our parameter space.
	\item Magnetic pressure can increase the size of the accretion disk.
\end{itemize}

When examining the full parameter space of initial conditions for the onset of gravitational collapse, we find that:
\begin{itemize}
	\item Our results for the size of the disk and the mass gain of the protostar scale with the initial mass of the cloud, despite the non-scalability of self-gravity and the thermodynamics considered.
	\item The thickness of the thick layer of the disk is controlled by the initial magnetic field strength.
	\item For a cloud with an initial density profile $\rho \propto r^{-1.5}$ and in solid-body rotation, the disk grows roughly linearly in size as $R_\text{disk} \approx [ 6380 M_\star/M_C - 98 ]\unit{au} $. The stellar mass grows approximately like $M_\star \propto (t/t_\text{ff})^{1.5..1.9}$.
	\item The size of the disk is more strongly determined by the initial distribution of density and rotational energy in the cloud than by the strength of the magnetic field.
	\item Multiple initial configurations of the cloud can produce a given disk size and (proto)stellar mass. This means that observations of disk-jet systems constrain (as opposed to determine) the possible conditions for the onset of gravitational collapse, and more measurements (distribution and strength of magnetic fields, for example) are needed to break the degeneracies.
\end{itemize}

In a follow-up article in preparation, we will perform a dynamical analysis of the magnetically-driven outflows of the same dataset we present here.

\begin{acknowledgements}
We thank Richard Nies for his contributions to the analysis of part of the dataset at the early stages of the project. GAO acknowledges financial support from the Deutscher Akademischer Austauschdienst (DAAD), under the program Research Grants - Doctoral Projects in Germany, and complementary financial support for the completion of the Doctoral degree by the University of Costa Rica, as part of their scholarship program for postgraduate studies in foreign institutions. RK acknowledges financial support via the Emmy Noether and Heisenberg Research Grants funded by the German Research Foundation (DFG) under grant no.~KU 2849/3 and 2849/9.
\end{acknowledgements}

\bibliographystyle{aa} 
\bibliography{jetsfld}

\begin{thebibliography}{50}
\expandafter\ifx\csname natexlab\endcsname\relax\def\natexlab#1{#1}\fi

\bibitem[{{Banerjee} \& {Pudritz}(2007)}]{Banerjee2007}
{Banerjee}, R. \& {Pudritz}, R.~E. 2007, \apj, 660, 479

\bibitem[{{Beltr{\'a}n} {et~al.}(2019){Beltr{\'a}n}, {Padovani}, {Girart},
  {Galli}, {Cesaroni}, {Paladino}, {Anglada}, {Estalella}, {Osorio}, {Rao},
  {S{\'a}nchez-Monge}, \& {Zhang}}]{Beltran2019}
{Beltr{\'a}n}, M.~T., {Padovani}, M., {Girart}, J.~M., {et~al.} 2019, \aap,
  630, A54

\bibitem[{{Beuther} {et~al.}(2018){Beuther}, {Mottram}, {Ahmadi}, {Bosco},
  {Linz}, {Henning}, {Klaassen}, {Winters}, {Maud}, {Kuiper}, {Semenov},
  {Gieser}, {Peters}, {Urquhart}, {Pudritz}, {Ragan}, {Feng}, {Keto},
  {Leurini}, {Cesaroni}, {Beltran}, {Palau}, {S{\'a}nchez-Monge},
  {Galvan-Madrid}, {Zhang}, {Schilke}, {Wyrowski}, {Johnston}, {Longmore},
  {Lumsden}, {Hoare}, {Menten}, \& {Csengeri}}]{Beuther2018}
{Beuther}, H., {Mottram}, J.~C., {Ahmadi}, A., {et~al.} 2018, \aap, 617, A100

\bibitem[{{Beuther} {et~al.}(2002){Beuther}, {Schilke}, {Menten}, {Motte},
  {Sridharan}, \& {Wyrowski}}]{2002ApJ...566..945B}
{Beuther}, H., {Schilke}, P., {Menten}, K.~M., {et~al.} 2002, \apj, 566, 945

\bibitem[{{Beuther} \& {Shepherd}(2005)}]{Beuther2005}
{Beuther}, H. \& {Shepherd}, D. 2005, in Astrophysics and Space Science
  Library, Vol. 324, Astrophysics and Space Science Library, ed. M.~S.~N.
  {Kumar}, M.~{Tafalla}, \& P.~{Caselli}, 105

\bibitem[{{Beuther} {et~al.}(2020){Beuther}, {Soler}, {Linz}, {Henning},
  {Gieser}, {Kuiper}, {Vlemmings}, {Hennebelle}, {Feng}, {Smith}, \&
  {Ahmadi}}]{Beuther2020}
{Beuther}, H., {Soler}, J.~D., {Linz}, H., {et~al.} 2020, \apj, 904, 168

\bibitem[{{Blandford} \& {Payne}(1982)}]{BlandfordPayne1982}
{Blandford}, R.~D. \& {Payne}, D.~G. 1982, \mnras, 199, 883

\bibitem[{{Carrasco-Gonz{\'a}lez} {et~al.}(2010){Carrasco-Gonz{\'a}lez},
  {Rodr{\'\i}guez}, {Anglada}, {Mart{\'\i}}, {Torrelles}, \&
  {Osorio}}]{Carrasco-Gonzalez2010}
{Carrasco-Gonz{\'a}lez}, C., {Rodr{\'\i}guez}, L.~F., {Anglada}, G., {et~al.}
  2010, Science, 330, 1209

\bibitem[{{Carrasco-Gonz{\'a}lez} {et~al.}(2021){Carrasco-Gonz{\'a}lez},
  {Sanna}, {Rodr{\'\i}guez-Kamenetzky}, {Moscadelli}, {Hoare}, {Torrelles},
  {Galv{\'a}n-Madrid}, \& {Izquierdo}}]{Carrasco-Gonzalez2021}
{Carrasco-Gonz{\'a}lez}, C., {Sanna}, A., {Rodr{\'\i}guez-Kamenetzky}, A.,
  {et~al.} 2021, \apjl, 914, L1

\bibitem[{{Ching} {et~al.}(2022){Ching}, {Li}, {Heiles}, {Li}, {Qian}, {Yue},
  {Tang}, \& {Jiao}}]{2022Natur.601...49C}
{Ching}, T.~C., {Li}, D., {Heiles}, C., {et~al.} 2022, \nat, 601, 49

\bibitem[{{Commer{\c{c}}on} {et~al.}(2022){Commer{\c{c}}on}, {Gonz{\'a}lez},
  {Mignon-Risse}, {Hennebelle}, \& {Vaytet}}]{Commercon2022}
{Commer{\c{c}}on}, B., {Gonz{\'a}lez}, M., {Mignon-Risse}, R., {Hennebelle},
  P., \& {Vaytet}, N. 2022, \aap, 658, A52

\bibitem[{{Commer{\c{c}}on} {et~al.}(2011){Commer{\c{c}}on}, {Teyssier},
  {Audit}, {Hennebelle}, \& {Chabrier}}]{Commercon2011rad}
{Commer{\c{c}}on}, B., {Teyssier}, R., {Audit}, E., {Hennebelle}, P., \&
  {Chabrier}, G. 2011, \aap, 529, A35

\bibitem[{{Galli} {et~al.}(2006){Galli}, {Lizano}, {Shu}, \&
  {Allen}}]{Galli2006}
{Galli}, D., {Lizano}, S., {Shu}, F.~H., \& {Allen}, A. 2006, \apj, 647, 374

\bibitem[{{Gieser} {et~al.}(2022){Gieser}, {Beuther}, {Semenov}, {Suri},
  {Soler}, {Linz}, {Syed}, {Henning}, {Feng}, {M{\"o}ller}, {Palau}, {Winters},
  {Beltr{\'a}n}, {Kuiper}, {Moscadelli}, {Klaassen}, {Urquhart}, {Peters},
  {Longmore}, {S{\'a}nchez-Monge}, {Galv{\'a}n-Madrid}, {Pudritz}, \&
  {Johnston}}]{2022AA...657A...3G}
{Gieser}, C., {Beuther}, H., {Semenov}, D., {et~al.} 2022, \aap, 657, A3

\bibitem[{{Girart} {et~al.}(2017){Girart}, {Estalella},
  {Fern{\'a}ndez-L{\'o}pez}, {Curiel}, {Frau}, {Galvan-Madrid}, {Rao},
  {Busquet}, \& {Ju{\'a}rez}}]{Girart2017}
{Girart}, J.~M., {Estalella}, R., {Fern{\'a}ndez-L{\'o}pez}, M., {et~al.} 2017,
  \apj, 847, 58

\bibitem[{{Girart} {et~al.}(2013){Girart}, {Frau}, {Zhang}, {Koch}, {Qiu},
  {Tang}, {Lai}, \& {Ho}}]{Girart2013}
{Girart}, J.~M., {Frau}, P., {Zhang}, Q., {et~al.} 2013, \apj, 772, 69

\bibitem[{{Goodman} {et~al.}(1993){Goodman}, {Benson}, {Fuller}, \&
  {Myers}}]{Goodman1993}
{Goodman}, A.~A., {Benson}, P.~J., {Fuller}, G.~A., \& {Myers}, P.~C. 1993,
  \apj, 406, 528

\bibitem[{{Guzm{\'a}n} {et~al.}(2010){Guzm{\'a}n}, {Garay}, \&
  {Brooks}}]{Guzman2010}
{Guzm{\'a}n}, A.~E., {Garay}, G., \& {Brooks}, K.~J. 2010, \apj, 725, 734

\bibitem[{{Hatchell} \& {van der Tak}(2003)}]{2003A&A...409..589H}
{Hatchell}, J. \& {van der Tak}, F.~F.~S. 2003, \aap, 409, 589

\bibitem[{{Hosokawa} \& {Omukai}(2009)}]{HosokawaOmukai2009evol}
{Hosokawa}, T. \& {Omukai}, K. 2009, \apj, 691, 823

\bibitem[{{Kee} \& {Kuiper}(2019)}]{Kee2019}
{Kee}, N.~D. \& {Kuiper}, R. 2019, \mnras, 483, 4893

\bibitem[{{K{\"o}lligan} \& {Kuiper}(2018)}]{Anders2018}
{K{\"o}lligan}, A. \& {Kuiper}, R. 2018, \aap, 620, A182

\bibitem[{{Kuiper} \& {Hosokawa}(2018)}]{KuiperHosokawa2018}
{Kuiper}, R. \& {Hosokawa}, T. 2018, \aap, 616, A101

\bibitem[{{Kuiper} {et~al.}(2010){Kuiper}, {Klahr}, {Beuther}, \&
  {Henning}}]{Kuiper2010circ}
{Kuiper}, R., {Klahr}, H., {Beuther}, H., \& {Henning}, T. 2010, \apj, 722,
  1556

\bibitem[{{Kuiper} {et~al.}(2011){Kuiper}, {Klahr}, {Beuther}, \&
  {Henning}}]{Kuiper2011}
{Kuiper}, R., {Klahr}, H., {Beuther}, H., \& {Henning}, T. 2011, \apj, 732, 20

\bibitem[{{Kuiper} {et~al.}(2020){Kuiper}, {Yorke}, \& {Mignone}}]{Kuiper2020}
{Kuiper}, R., {Yorke}, H.~W., \& {Mignone}, A. 2020, \apjs, 250, 13

\bibitem[{{Laor} \& {Draine}(1993)}]{LaorDraine1993}
{Laor}, A. \& {Draine}, B.~T. 1993, \apj, 402, 441

\bibitem[{{Lynden-Bell}(2003)}]{Lynden-Bell2003}
{Lynden-Bell}, D. 2003, \mnras, 341, 1360

\bibitem[{{Machida} \& {Hosokawa}(2020)}]{Machida2020}
{Machida}, M.~N. \& {Hosokawa}, T. 2020, \mnras, 499, 4490

\bibitem[{{Machida} {et~al.}(2007){Machida}, {Inutsuka}, \&
  {Matsumoto}}]{Machida2007}
{Machida}, M.~N., {Inutsuka}, S.-i., \& {Matsumoto}, T. 2007, \apj, 670, 1198

\bibitem[{{Marchand} {et~al.}(2022){Marchand}, {Guillet}, {Lebreuilly}, \& {Mac
  Low}}]{Marchand2022}
{Marchand}, P., {Guillet}, V., {Lebreuilly}, U., \& {Mac Low}, M.-M. 2022,
  arXiv e-prints, arXiv:2202.11625

\bibitem[{{Marchand} {et~al.}(2016){Marchand}, {Masson}, {Chabrier},
  {Hennebelle}, {Commer{\c{c}}on}, \& {Vaytet}}]{Marchand2016}
{Marchand}, P., {Masson}, J., {Chabrier}, G., {et~al.} 2016, \aap, 592, A18

\bibitem[{{Masson} {et~al.}(2016){Masson}, {Chabrier}, {Hennebelle}, {Vaytet},
  \& {Commer{\c{c}}on}}]{Masson2016}
{Masson}, J., {Chabrier}, G., {Hennebelle}, P., {Vaytet}, N., \&
  {Commer{\c{c}}on}, B. 2016, \aap, 587, A32

\bibitem[{{Matsushita} {et~al.}(2017){Matsushita}, {Machida}, {Sakurai}, \&
  {Hosokawa}}]{Matsushita2017}
{Matsushita}, Y., {Machida}, M.~N., {Sakurai}, Y., \& {Hosokawa}, T. 2017,
  \mnras, 470, 1026

\bibitem[{{Mignon-Risse} {et~al.}(2021){Mignon-Risse}, {Gonz{\'a}lez},
  {Commer{\c{c}}on}, \& {Rosdahl}}]{Mignon-Risse2021}
{Mignon-Risse}, R., {Gonz{\'a}lez}, M., {Commer{\c{c}}on}, B., \& {Rosdahl}, J.
  2021, \aap, 652, A69

\bibitem[{{Mignone} {et~al.}(2007){Mignone}, {Bodo}, {Massaglia}, {Matsakos},
  {Tesileanu}, {Zanni}, \& {Ferrari}}]{Mignone2007}
{Mignone}, A., {Bodo}, G., {Massaglia}, S., {et~al.} 2007, \apjs, 170, 228

\bibitem[{{Moscadelli} {et~al.}(2021){Moscadelli}, {Beuther}, {Ahmadi},
  {Gieser}, {Massi}, {Cesaroni}, {S{\'a}nchez-Monge}, {Bacciotti},
  {Beltr{\'a}n}, {Csengeri}, {Galv{\'a}n-Madrid}, {Henning}, {Klaassen},
  {Kuiper}, {Leurini}, {Longmore}, {Maud}, {M{\"o}ller}, {Palau}, {Peters},
  {Pudritz}, {Sanna}, {Semenov}, {Urquhart}, {Winters}, \&
  {Zinnecker}}]{Moscadelli2021multi}
{Moscadelli}, L., {Beuther}, H., {Ahmadi}, A., {et~al.} 2021, \aap, 647, A114

\bibitem[{{Moscadelli} {et~al.}(2022){Moscadelli}, {Sanna}, {Beuther}, {Oliva},
  \& {Kuiper}}]{Moscadelli2022}
{Moscadelli}, L., {Sanna}, A., {Beuther}, H., {Oliva}, A., \& {Kuiper}, R.
  2022, Nat. Astron., https://doi.org/10.1038/s41550-022-01754-4

\bibitem[{{Mouschovias} \& {Spitzer}(1976)}]{MouschoviasSpitzer1976}
{Mouschovias}, T.~C. \& {Spitzer}, L., J. 1976, \apj, 210, 326

\bibitem[{{Mueller} {et~al.}(2002){Mueller}, {Shirley}, {Evans}, \&
  {Jacobson}}]{2002ApJS..143..469M}
{Mueller}, K.~E., {Shirley}, Y.~L., {Evans}, Neal~J., I., \& {Jacobson}, H.~R.
  2002, \apjs, 143, 469

\bibitem[{{Myers} {et~al.}(2013){Myers}, {McKee}, {Cunningham}, {Klein}, \&
  {Krumholz}}]{Myers2013}
{Myers}, A.~T., {McKee}, C.~F., {Cunningham}, A.~J., {Klein}, R.~I., \&
  {Krumholz}, M.~R. 2013, \apj, 766, 97

\bibitem[{{Nakano} {et~al.}(2002){Nakano}, {Nishi}, \&
  {Umebayashi}}]{Nakano2002}
{Nakano}, T., {Nishi}, R., \& {Umebayashi}, T. 2002, \apj, 573, 199

\bibitem[{{Oliva} \& {Kuiper}(2020)}]{Oliva2020}
{Oliva}, G.~A. \& {Kuiper}, R. 2020, \aap, 644, A41

\bibitem[{{Purser} {et~al.}(2016){Purser}, {Lumsden}, {Hoare}, {Urquhart},
  {Cunningham}, {Purcell}, {Brooks}, {Garay}, {G{\'u}zman}, \&
  {Voronkov}}]{Purser2016}
{Purser}, S.~J.~D., {Lumsden}, S.~L., {Hoare}, M.~G., {et~al.} 2016, \mnras,
  460, 1039

\bibitem[{{Rodr{\'\i}guez-Kamenetzky}
  {et~al.}(2017){Rodr{\'\i}guez-Kamenetzky}, {Carrasco-Gonz{\'a}lez}, {Araudo},
  {Romero}, {Torrelles}, {Rodr{\'\i}guez}, {Anglada}, {Mart{\'\i}}, {Perucho},
  \& {Valotto}}]{Rodriguezkamenetzky2017}
{Rodr{\'\i}guez-Kamenetzky}, A., {Carrasco-Gonz{\'a}lez}, C., {Araudo}, A.,
  {et~al.} 2017, \apj, 851, 16

\bibitem[{{Rosen} \& {Krumholz}(2020)}]{Rosen2020}
{Rosen}, A.~L. \& {Krumholz}, M.~R. 2020, \aj, 160, 78

\bibitem[{{Seifried} {et~al.}(2011){Seifried}, {Banerjee}, {Klessen}, {Duffin},
  \& {Pudritz}}]{Seifried2011}
{Seifried}, D., {Banerjee}, R., {Klessen}, R.~S., {Duffin}, D., \& {Pudritz},
  R.~E. 2011, \mnras, 417, 1054

\bibitem[{{Shakura} \& {Sunyaev}(1973)}]{ShakuraSunyaev1973}
{Shakura}, N.~I. \& {Sunyaev}, R.~A. 1973, \aap, 24, 337

\bibitem[{{Troland} \& {Crutcher}(2008)}]{2008ApJ...680..457T}
{Troland}, T.~H. \& {Crutcher}, R.~M. 2008, \apj, 680, 457

\bibitem[{{Tsukamoto} {et~al.}(2021){Tsukamoto}, {Machida}, \&
  {Inutsuka}}]{Tsukamoto2020}
{Tsukamoto}, Y., {Machida}, M.~N., \& {Inutsuka}, S. 2021, \apj, 913, 148

\end{thebibliography}

\appendix

\section{Numerical convergence} \label{s:convergence}
\begin{figure*}
\includegraphics[width=\textwidth]{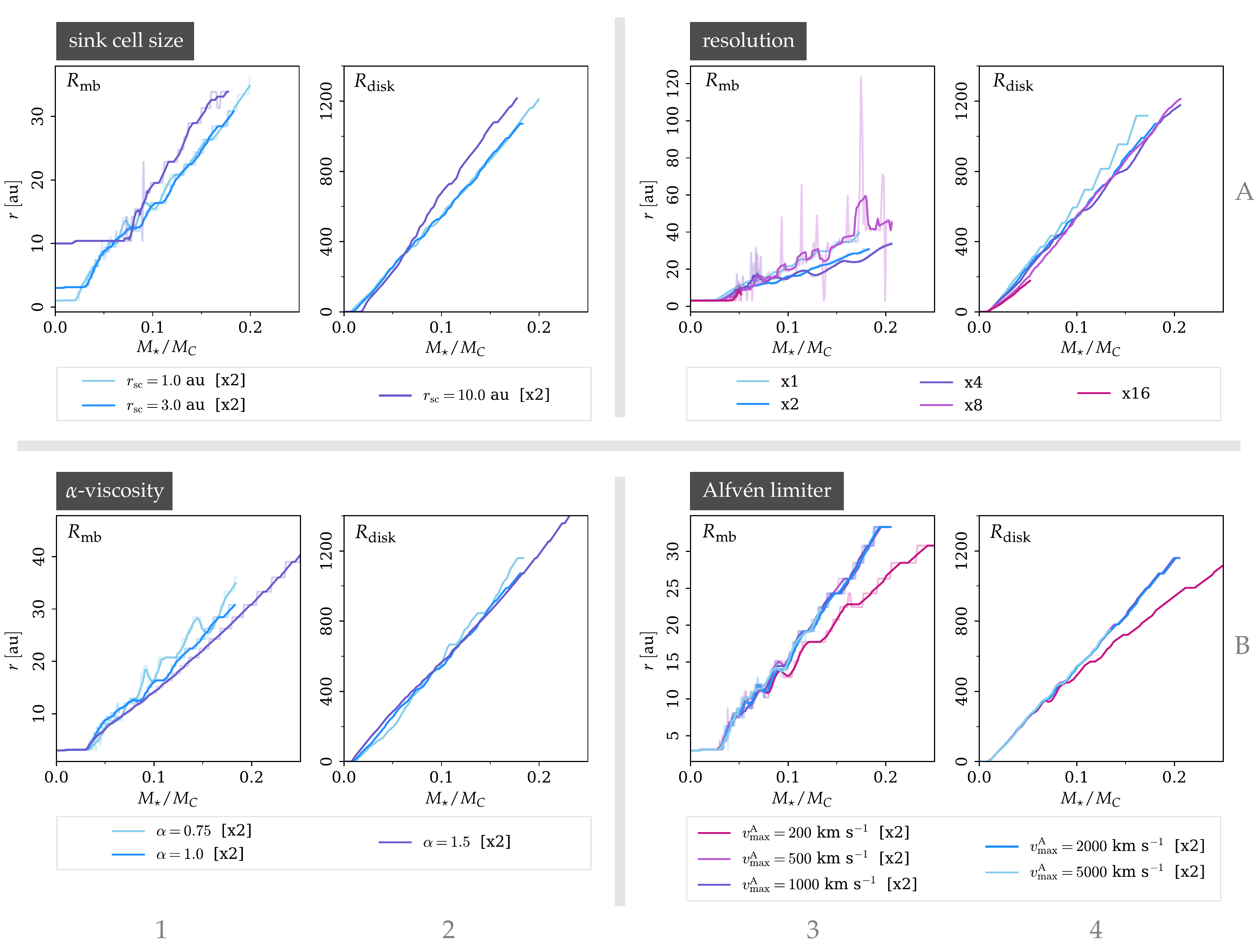}
\caption{Variation of the radius of the disk and the magnetic braking radius for different the inner boundary (sink cell size), resolution of the grid, the $\alpha$-viscosity used to model gravitational torques and the Alfvén limiter.}
\label{rdisk_num}
\end{figure*}

As part of the parameter scan performed for this study, we investigated the effects of the grid and other numerically-relevant parameters on our findings. Here, we focus on checking the effect on the extent of the disk, i.e., its inner and outer radii, which we present in Fig. \ref{rdisk_num}.

\subsection{Sink cell size}
We tried sink cell radii of $r_\mathrm{sc} = 1, 3, \text{ and } 10 \unit{au}$ as our inner boundary. As shown in panels A1 and A2 of Fig. \ref{rdisk_num}, the results for the radius of the disk and the magnetic braking radius are very consistent for $r_\mathrm{sc}=1$ and $3\unit{au}$, but larger inner and outer radii at late phases of the simulation for $r_\mathrm{sc} = 10\unit{au}$. Of special interest is the fact that for the smaller sink sizes, both the size and the time of formation of the magnetically-braked region region does not change, which supports the idea that the effect is physical and not due to the inner boundary conditions.

During the formation of the massive star starting from the $100\unit{M_\odot}$ cloud, the stellar radius computed with the stellar evolutionary tracks increases up to a maximum of 150 solar radii, which correspond to approx. 0.7 astronomical units. The use of a smaller sink cell would not be accurate because it would require the solution of the full set of stellar structure equations, which we do not do. With this in mind, we conclude that a sink cell size of $1\unit{au}$ is close to the reasonable limit for the approximations considered in this study, and a sink cell size of $3\unit{au}$ is enough to replicate the results of the smaller sink.

\subsection{Resolution}
We conducted the study of the fiducial case in five different resolutions, each one doubling the amount of cells in each one of the directions (radial and polar). The simulation x16 was only run until $t=13.45\unit{kyr}$ due to the high computational cost. The  radius of the disk is very similar in all of the simulations. The magnetic braking radius appears somewhat later in the high resolution simulations and it shows more variability for the grid x8. This could be due to the increased resolution of the effects of the outflows, which will be discussed in \PaperII. However, during the comparison of the disk radii, it is important to consider that our definition of the size of the disk, based on a simple threshold to Keplerianity, is affected by the size of the cells, rounding to the nearest cell that satisfies the criterion.

\subsection{Viscosity}
The $\alpha$-viscosity model to gravitational torques produced by spiral arms in the self-gravitating disk introduces an additional parameter. For this reason, we investigated what happens if the value of $\alpha$ is changed to $0.75$ and $1.5$ from its fiducial value of $1$. Panels B1 and B2 in Fig. \ref{rdisk_num} show relatively consistent results in all cases, even though with some local differences. The magnetic braking region appears when the star is of the same mass in all cases, and the radius of the disk averages approximately to the same value over time. Even though viscosity transports angular momentum, the qualitative behavior is different than the effect of resistivity and initial magnetic field, both of which significantly alter the formation time of the magnetically-braked region and its size. This marks the difference between magnetic braking and angular momentum transport by viscosity.

\subsection{Alfvén limiter}
We employed an upper limit to the Alfvén speed
\begin{equation}
	v_A = \frac{B}{\sqrt{4\pi\rho}},
\end{equation}
by introducing a locally-variable density floor such that the Alfvén speed never exceeds the limit, which is defined a priori. This limiter is introduced to avoid the Alfvén speed to increase indefinitely. The time step remains within limits that allow us to cover the desired timescales in a reasonable computing time. This approach has the disadvantage that artificial mass is created in the process, which in theory could affect the dynamics of the gas. A more detailed discussion of the effects of the Alfvén limiter will be offered in \PaperII on jet physics, however, we mention here that with the values chosen for the simulation series x4 and x8, $2000$ and $1000\unit{km\,s^{-1}}$ respectively, artificial mass below $0.002\unit{M_\sun}$ is generated at the end of the simulation. This value is very small compared to the mass of the cloud and should not change the dynamics of the disk and the outflows. In turn, for simulation series x2, an Alfvén limiter of $2000\unit{km\,s^{-1}}$ produces an additional $0.2\unit{M_\odot}$ of artificial mass, with lower values yielding over $1\unit{M_\odot}$, i.e., more than one per cent of the initial mass of the cloud. With these values of the limiter and the artificial mass in mind, we examine the extent of the disk in panels B3 and B4 of Fig. \ref{rdisk_num} as a probe for effects in the dynamics of the disk and find that only in the case of an exceptionally low Alfvén limiter of $200\unit{km\,s^{-1}}$ there are appreciable changes to the inner and outer radii of the disk for late times.

\end{document}